\documentclass[aps,prd,floatfix,preprintnumbers,altaffilletter,superscriptaddress,reprint,longbibliography,twocolumn]{revtex4-1}

\usepackage{amssymb}
\usepackage{amsmath}
\usepackage{aas_macros}
\usepackage[dvipsnames]{xcolor}
\usepackage[colorlinks,urlcolor=NavyBlue,citecolor=NavyBlue,linkcolor=NavyBlue]{hyperref}
\usepackage{color,units}
\usepackage{xspace}
\usepackage{subfigure}
\usepackage{comment}
\usepackage{bm}
\usepackage{dcolumn}
\usepackage{hyperref}
\usepackage{enumitem} 
\usepackage{array}
\usepackage{nicematrix}
\usepackage{hhline}
\usepackage{graphicx}
\usepackage{multirow}
\usepackage[normalem]{ulem}
\usepackage{soul}
\usepackage[math]{cellspace}
\usepackage{nicematrix} 
\usepackage{makecell}
\usepackage{acronym}
\usepackage{caption}
\usepackage{tikz}
\usetikzlibrary{arrows.meta, positioning, fit, backgrounds, calc}

\begin{document}

\title{Fortifying gravitational-wave population inference with normalizing flows}

\author{Christian Adamcewicz}
\affiliation{School of Physics and Astronomy, Monash University, VIC 3800, Australia}
\affiliation{OzGrav: The ARC Centre of Excellence for Gravitational-Wave Discovery, Clayton, VIC 3800, Australia}

\author{Hugh McDougall}
\affiliation{School of Mathematics and Physics, University of Queensland, St Lucia, QLD 4072, Australia}
\affiliation{OzGrav: The ARC Centre of Excellence for Gravitational-Wave Discovery, St Lucia, QLD 4072, Australia}

\author{Paul D. Lasky}
\author{Eric Thrane}
\affiliation{School of Physics and Astronomy, Monash University, VIC 3800, Australia}
\affiliation{OzGrav: The ARC Centre of Excellence for Gravitational-Wave Discovery, Clayton, VIC 3800, Australia}

\begin{abstract}
As the LIGO-Virgo-KAGRA collaboration's (LVK's) gravitational-wave transient catalog grows, we are learning a wealth of information from the population properties of binary black hole mergers.
Events in the catalog are represented with posterior samples describing the astrophysical parameters for each event.
Population studies combine these samples to measure the distribution of astrophysical parameters such as black hole masses and spins.
However, the posterior-sample representation of each event is only approximate.
We construct a mock population with masses drawn from an astrophysically-motivated distribution with sharp features.
Using this, we demonstrate that when $\gtrsim 300$ events are combined, even with each event's posterior represented by $1 \times 10^4 {-} 2 \times 10^4$ samples, the numerical error can become large enough that the resulting population inference is unreliable.
We consider two solutions.
In the short term, we show that \textit{nested samples} (already produced by LVK analyses) can be used to more accurately describe each event in population studies.
But this will only grant a temporary reprieve until the nested-sample representation becomes inadequate.
In the longer term, we propose to represent each event with a \textit{normalizing flow}.
In order to represent each event with sufficient accuracy, each normalizing flow can be used to generate an arbitrarily large number of new posterior samples with a significantly reduced computational cost relative to traditional sampling methods.
When compared to nested sampling, our normalizing flows produce posterior draws with a median of $\approx 80\%$ fewer likelihood evaluations per sample, while also providing greater opportunity for parallelization.
We believe refinement of normalizing flow architectures and training techniques in future works could further reduce this per-sample cost significantly.
\end{abstract}

\maketitle

\section{Introduction}
Fueled by advancements in detection capabilities \citep{Capote:2024rmo, LIGO:2024kkz, aLIGO:2020wna, LIGOO4Detector:2023wmz, membersoftheLIGOScientific:2024elc}, the recent release of the LIGO-Virgo-KAGRA collaboration's (LVK's; \citep{LIGOScientific:2014pky, VIRGO:2014yos, KAGRA:2020tym}) Gravitational-Wave Transient Catalog 5 (GWTC-5; \citep{LIGOScientific:2026sit, LIGOScientific:2026ifv, LIGOScientific:2026wfs, LIGOScientific:2026jgl}) has brought the total number of observed compact binary mergers to 267.
With this growing catalog, there have been a wide range of analyses characterizing the population of gravitational-wave binaries.
These include fitting the distributions of source properties like black hole masses and spins (e.g., \citep{LIGOScientific:2026ctl, LIGOScientific:2025pvj, Callister:2024cdx}), cosmological measurements that rely on the distance to each source (e.g., \citep{LIGOScientific:2026uyd, LIGOScientific:2025jau, Gair:2022zsa}), and tests of general relativity (e.g., \citep{LIGOScientific:2026qni, LIGOScientific:2026fcf}) to name a few.

Typically, in order to infer population-level properties across the catalog, we begin with \textit{posterior samples}: random draws from the posterior distribution for the 15 parameters that characterize each binary.
These fiducial samples are produced using simple, fixed prior distributions.
During hierarchical inference, the likelihood is computed by importance-sampling these fiducial posterior samples.
Importance sampling allows us to reuse the fiducial prior (known as the ``proposal distribution'') in order to obtain a result for the population prior (known as the ``target distribution''; see Ref.~\citep{Thrane:2018qnx} for example).
In doing so, we approximate a continuous, 15-dimensional likelihood surface with a finite number of posterior samples.

This approximation introduces a numerical error.
While typically small for a single event, when many events are combined, the total error can become large enough to significantly bias inferences \citep{Heinzel:2025ogf, Talbot:2023pex, Essick:2022ojx, Farr:2019rap}.
Such sampling errors may already affect some LVK analyses.
The current approach is to simply mark regions of parameter space where the inference becomes unreliable; see, e.g., Refs.~\citep{Talbot:2023pex, Essick:2022ojx, Farr:2019rap}.
Of course, this is not a solution so much as it is an acknowledgment of a problem.
As additional events are added to the catalog, the region of parameter space where we can carry out reliable inference will shrink and eventually vanish.
The extent of the looming problem is brought into relief when we consider the fact that next-generation observatories are expected to detect ${\cal O}(10^5)$ gravitational-wave signals every year, increasing the size of our current catalogs by orders of magnitude \cite{Reitze:2019iox}.

In this work, we exemplify the problem of numerical bias in population studies from posterior sample representations of gravitational-wave events and propose two solutions.
The remainder of this work is structured as follows.
In Section~\ref{sec:motivation} we motivate the study with a mathematical description of numerical bias in population studies.
Next, we demonstrate the problem using simulated data in Section~\ref{sec:mock_problem}.
In Section~\ref{sec:easy_fixes}, we describe simple methods for decreasing numerical error that can easily be implemented with existing data products.
Using our synthetic data, we demonstrate that these improvements---while potentially sufficient for current population analyses---may not be adequate to ensure bias-free results with catalogs of $\gtrsim 300$ events.
In Section~\ref{sec:flows}, we propose the use of normalizing flows as a longer-term solution to control numerical errors in population studies.
We demonstrate the plausibility of this by building, training and utilizing simple flows to solve our mock population problem.
We conclude in Section~\ref{sec:discussion} with a discussion of our findings and future work.

\section{Motivation}\label{sec:motivation}
We begin by assuming a model for the distribution of some set of compact binary parameters $\theta$ (e.g., masses, spins, redshifts, etc.).
The shape of this distribution depends on model hyper-parameters $\Lambda$.
The population model is a conditional prior, which we denote as: $\pi(\theta|\Lambda)$.
Our goal is then to infer the hyper-posterior distribution $p(\Lambda|d)$ using gravitational-wave data $d$ from the observed population of merging binaries.

To do so, we define the population likelihood
\begin{equation} \label{eq:population_likelihood}
    \mathcal{L}_\mathrm{pop}(d|\Lambda) = \prod_i^N \int d\theta \ \mathcal{L}(d_i|\theta) \pi(\theta|\Lambda),
\end{equation}
where the product is over $N$ events, and $\mathcal{L}(d_i|\theta)$ is the likelihood of the data from event $i$ given binary parameters $\theta$.
For the sake of simplicity, we ignore selection effects for now, but it would not affect our point about numerical error stemming from stacked posteriors if we included them.
We elaborate on this in Section~\ref{sec:discussion}.
Our worked examples in Sections~\ref{sec:mock_problem}-\ref{sec:flows} do include selection effects.

Due to the high-dimensionality of $\theta$, we typically rely on Monte-Carlo integration to evaluate this population likelihood.
Specifically, we take $n$ samples of $\theta$ for each event, drawn from the fiducial single-event posteriors 
\begin{align}
p_{\O}(\theta|d_i) \propto \mathcal{L}(d_i|\theta)\pi_{\O}(\theta) ,
\end{align}
where $\pi_{\O}(\theta)$ is the fiducial prior used for initial parameter estimation (PE).
We then reweight each sample to follow the population model described by hyper-parameters $\Lambda$.
The weights are given  by
\begin{equation} \label{eq:population_weights}
    w(\theta_i^k|\Lambda) = \frac{\pi(\theta_i^k|\Lambda)}{\pi_{\O}(\theta_i^k)}.
\end{equation}
Here, $\theta_i^k$ is the $k^\mathrm{th}$ sample for event $i$.
The population likelihood in Eq.~\ref{eq:population_likelihood} can then be approximated as
\begin{equation} \label{eq:population_likelihood_mc}
    \mathcal{L}_\mathrm{pop}(d|\Lambda) \approx \prod_i^N \mathcal{Z}_{\O}(d_i) E_{n}\left[ w(\theta_i^k|\Lambda)\right],
\end{equation}
where $\mathcal{Z}_{\O}(d_i)$ is the Bayesian evidence for event $i$ obtained during fiducial sampling.
Meanwhile, $E_{n}\left[w(\theta_i^k|\Lambda)\right]$ is shorthand for the expectation value of $w(\theta_i^k|\Lambda)$ calculated with $n$ samples.

By approximating the integrals in the population likelihood with $n$ samples, we introduce a numerical error, which is most commonly (but, see also Refs.~\cite{Essick:2022ojx, Farr:2019rap}) quantified by the variance:
\begin{equation} \label{eq:population_uncertainty}
    \sigma_\mathcal{L}^2 = \sum_i^N \frac{1}{n}\left(E_{n}\left[ w(\theta_i^k|\Lambda)^2\right] - E_{n}\left[ w(\theta_i^k|\Lambda)\right]^2\right).
\end{equation}
If sufficiently large, this variance can bias population inferences \cite{Heinzel:2025ogf, Talbot:2023pex}, and even result in numerical instabilities while sampling.
To avoid this, researchers typically set some threshold on the variance, below which the uncertainty introduced by Monte-Carlo integration is negligible.
If, during sampling, this threshold is exceeded, the associated value of $\Lambda$ is discarded and/or marked as an unreliable region of the hyper-parameter space.
Typically, this threshold is set at a value of $\sigma_\mathcal{L}^2=1$ \cite{Talbot:2023pex} (although, this may be overly strict in some cases \cite{Heinzel:2025ogf}).

In some instances, enforcing $\sigma_\mathcal{L}^2$ is below threshold can a~priori rule out areas of interest in the hyper-posterior, rendering certain hypotheses untestable.
Worse, if these effects are not carefully monitored, we can be misled to believe that features in the hyper-posterior are driven by the data, when they are in fact being driven by uncertainty-based prior cuts.

We consider three factors affecting the size of the uncertainty in Eq.~\ref{eq:population_uncertainty}:
\begin{enumerate}
    \item Increasing variance between the weights $w(\theta_i^k|\Lambda)$ increases the uncertainty,
    \item Increasing the number of events $N$ increases the uncertainty, and
    \item Increasing the number of fiducial samples $n$ decreases the uncertainty.
\end{enumerate}
In practice, we have no control over the variance in weights (item 1), which becomes more problematic in models with sharp features.
In theory, one could perform parameter estimation with a fiducial prior $\pi_{\O}(\theta)$ that is chosen to closely (or exactly) resemble the population model $\pi(\theta|\Lambda)$, so as to minimize the variance between weights (see, e.g., Refs.~\citep{Adamcewicz:2025phm, Adamcewicz:2023szp, Tong:2022iws, Galaudage:2021rkt}).
However, regularly repeating parameter estimation on an entire catalog to maximize performance for a single population model is time consuming and computationally prohibitive.

Meanwhile, as the size of the catalog grows, we invariably increase the uncertainty due to Monte-Carlo integration in the population likelihood (item 2).
If we assume that individual events' contributions to the likelihood variance do not change notably over time, Eq.~\ref{eq:population_uncertainty} suggests that $\sigma_\mathcal{L}^2$ should increase approximately linearly with the number of events.
Ref.~\citep{Talbot:2023pex} finds experimental support for this scaling by analyzing several mock catalogs of varying sizes and measuring the corresponding trend in the variance.
On the other hand, Eq.~\ref{eq:population_uncertainty} implies that the likelihood variance is \textit{inversely} proportional to the number of samples $n$ used for reweighting (item 3).
This is again evidenced experimentally using mock catalogs in Ref.~\citep{Talbot:2023pex}.

Following this, the solutions explored in this paper work by increasing the number of effective samples $n$ representing each event in order to control the likelihood variance.
From the above, we might infer that such solutions will remain effective so long as we are able to follow each increase in catalog size with a linearly proportional increase in per-event sample size.
With this being said, we do not to explore the scaling here, but rather, show that this solution is effective for current and near-future catalogs (of up to 500 events).

Regardless of the exact proportionalities between likelihood variance $\sigma_\mathcal{L}^2$, catalog size $N$, and number of samples per event $n$, so long as some scaling exists (see discussions in Refs.~\citep{Heinzel:2025ogf, Talbot:2023pex, Essick:2022ojx}), this method will not sustain bias-free population inference indefinitely.
Thus, we argue below (see Section~\ref{sec:discussion} in particular) that the long-term solution to this problem is to develop sample-free representations of single-event likelihoods that can gradually be improved as the catalog grows, providing the equivalent performance of an adequate number of posterior samples.
The tools we develop in this work may form an integral part of this process.

\section{Demonstration}\label{sec:mock_problem}
To illustrate this problem further and test potential solutions, we construct a mock catalog of 300 binary black hole events.
We also extend our demonstration to 500 mock events in Appendix~\ref{sec:500_event}.
We focus on the 300 event case here, as this catalog size is more immediately relevant given current number of binary black hole events in GWTC-5, being 259 \citep{LIGOScientific:2026wfs, LIGOScientific:2026ctl}.
Black hole spins in our mock catalog follow the default ``strongly-parameterized'' distributions used in Refs.~\cite{LIGOScientific:2026ctl, LIGOScientific:2025pvj}.
Thus, spin magnitudes are drawn from independent and identically distributed normal distributions, truncated so that $\chi$ is between 0 and 1:
\begin{equation}
    \pi_\chi(\chi_1,\chi_2|\Lambda) =
    \mathcal{N}_{[0,1]}(\chi_1|\mu_\chi,\sigma_\chi)
    \mathcal{N}_{[0,1]}(\chi_2|\mu_\chi,\sigma_\chi),
\end{equation}
and cosine spin tilts are drawn from a uniform and Gaussian mixture model \citep{Talbot:2017yur, Vitale:2022dpa}:
\begin{multline}
    \pi_t(\cos t_1,\cos t_2|\Lambda) = \\
    \xi_t \mathcal{N}_{[-1,1]}(\cos t_1|\mu_t,\sigma_t) \mathcal{N}_{[-1,1]}(\cos t_2|\mu_t,\sigma_t) + \\
    (1-\xi_t) \mathcal{U}_{[-1,1]}(\cos t_1) \mathcal{U}_{[-1,1]}(\cos t_2).
\end{multline}
Here, $\mathcal{N}_{[a,b]}(x|\mu,\sigma)$ denotes a normal distribution with mean $\mu$ and width $\sigma$, truncated to $x \in [a,b]$, and $\mathcal{U}_{[a,b]}(x)$ indicates a uniform distribution over $x \in [a,b]$.

Meanwhile, redshifts are distributed according to the default ``strongly-parameterized'' power-law model from Refs.~\cite{LIGOScientific:2026ctl, LIGOScientific:2025pvj} (see Eq.~B25 therein; see also Ref.~\citep{Fishbach:2018edt}).
Masses in our mock catalog are drawn from a distribution $\pi_m(m_1,m_2|\Lambda)$ consisting of a power-law plus two peaks at $9 M_\odot$ and $30 M_\odot$, and an empty gap from $10M_\odot {-}15 M_\odot$.
For a mathematical definition of $\pi_m(m_1,m_2|\Lambda)$, see Eq.~2 from Ref.~\cite{Adamcewicz:2024jkr}.
This model is astrophysically motivated by Refs.~\cite{Legred:2026oiz, Adamcewicz:2024jkr, Galaudage:2024meo, Schneider:2023mxe, Schneider:2020vvh}, and is chosen because it includes a sharp gap and narrow peaks.
Models with sharp or narrow features tend to produce large variances in weights $w(\theta_i^k|\Lambda)$ for nearby events, thus large variances in the population likelihood $\sigma_\mathcal{L}^2$.
However, in this particular example, we found variance cuts more drastically affected inference of the spin distribution.
Therefore, while we do fit the distributions of component masses and redshift in our demonstrations, for brevity, we do not discuss them henceforth, and focus on the spin distribution.

We keep draws with a network matched-filter SNR of $\geq 10$ when their associated signal is injected into Gaussian noise representative of a two-detector network consisting of LIGO Hanford and Livingston operating at design sensitivity \citep{Romero-Shaw:2020owr, KAGRA:2013rdx}.
We also construct an associated set of $2.5 \times 10^{7}$ ``found injections'' based on the detection threshold to counter selection effects during population inference (see Refs.~\citep{Farr:2019rap, Thrane:2018qnx, Mandel:2018mve, Tiwari:2017ndi} for more information).
We make this injection set large enough so that the reweighting uncertainty is dominated by the importance sampling of events in the catalog (e.g., by Eq.~\ref{eq:population_uncertainty}; see Section~\ref{sec:discussion} for more discussion).

Relying on the \texttt{IMRPhenomXP} waveform model \citep{Pratten:2020ceb}, we inject signals for these mock events into Gaussian noise and perform fiducial parameter estimation on the resulting strain data using \textsc{Bilby} \citep{Romero-Shaw:2020owr, Ashton:2018jfp}.
To improve convergence, we explicitly marginalize over distance and time; see, e.g., Ref.~\cite{Thrane:2018qnx}.
We have chosen a waveform model with no higher-order modes, so that we can also analytically marginalize over phase \citep{Veitch:2014wba}.
We do so for the benefit of our normalizing flow approximations in Section~\ref{sec:flows}, as phase is difficult to capture using flow representations \citep{Dax:2022pxd}.
Ref.~\cite{Dax:2022pxd} demonstrates a method to avoid fitting phase in normalizing flows while retaining higher-order modes (effectively marginalizing over phase numerically rather than analytically; see also Ref.~\cite{Roulet:2024hwz}), but we choose to avoid higher-order modes entirely to keep our demonstration simple.

For our parameter estimation, we use the nested sampler \textsc{Dynesty} \citep{Speagle:2019ivv} along with sampler settings typical of LVK production-level analyses \citep{LIGOScientific:2026wfs, LIGOScientific:2025slb}.
This produces between $\approx 1\times10^4{-}2\times10^4$ samples (with a median of $\approx 1.3\times10^{4}$ samples) representing the fiducial posterior $p_{\O}(\theta|d_i)$ for each mock event.

\begin{figure*}
    \includegraphics[width=0.85\textwidth]{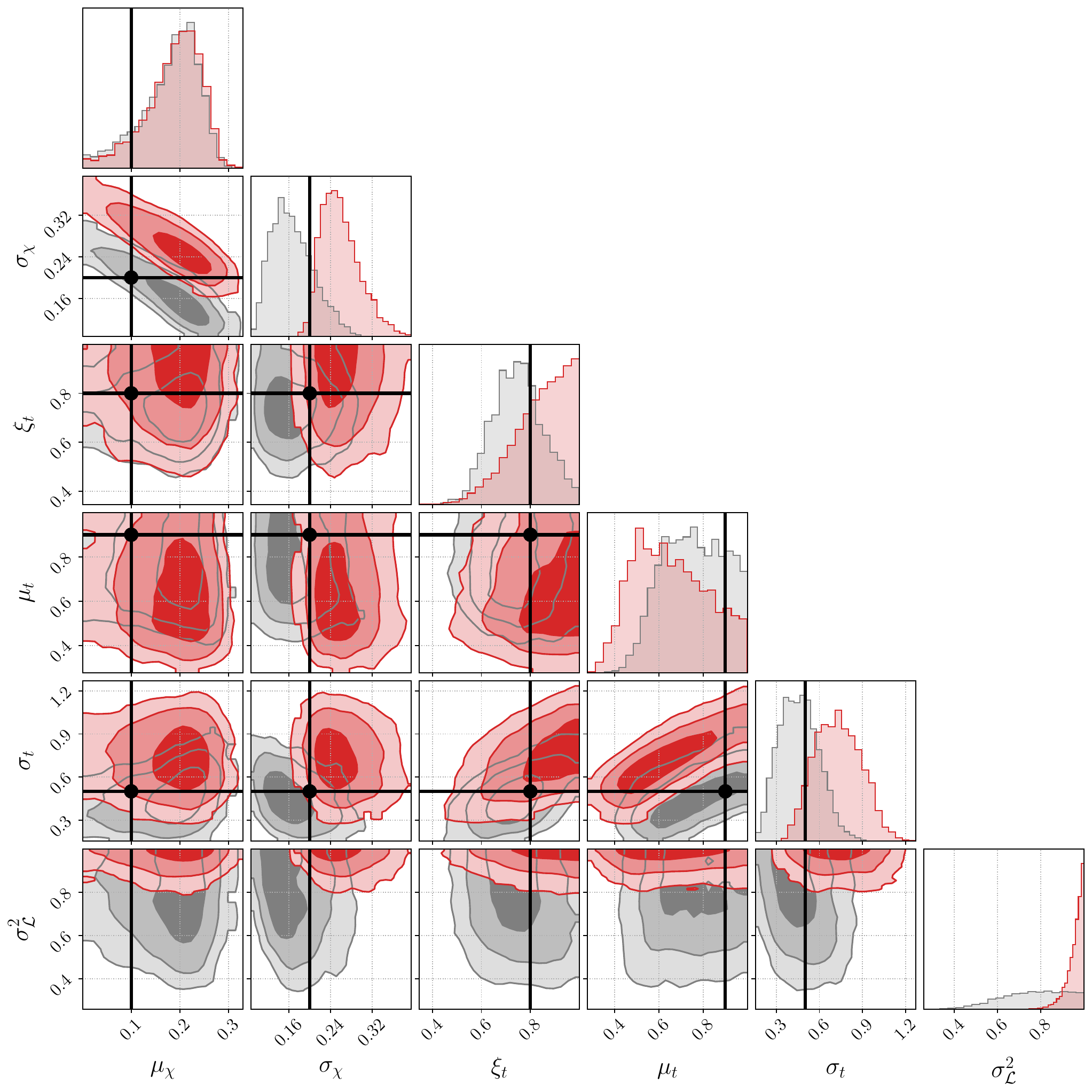}
    \caption{
    Posteriors for hyper-parameters governing the spin distribution, along with the distribution of likelihood variances $\sigma_\mathcal{L}^2$ associated with the hyper-posterior samples.
    The gray curves are inferred using $\approx 1\times10^4{-}2\times10^4$ fiducial samples, while the red curves are inferred using $2 \times 10^3$.
    From lightest to darkest, the shades in two-dimensional panels indicate 99\%, 90\%, and 50\% credible regions.
    The true injected value is shown in black.
    We see significant differences between the two posteriors, with the red contours consistently excluding the injected value.
    Both results have a substantial number of hyper-posterior samples near the $\sigma_\mathcal{L}^2 = 1$ threshold, indicating that a significant portion the hyper-posterior is being excluded.
    Clearly, the $2 \times 10^3$ posterior samples are not enough for reliable hierarchical inference in this example.
    As we will see later, the posterior inferred with all $\approx 1\times10^4{-}2\times10^4$ samples is also misleading as a result of variance cuts, albeit to a lesser degree.
    }
    \label{fig:original_gap_posterior}
\end{figure*}

Representing each event with all $\approx 1\times10^4{-}2\times10^4$ of these posterior samples, we perform a population analysis -- fitting $\Lambda$ for the same model we used to generate the data.
We carry out this fit with \textsc{GWPopulation} \citep{Talbot:2024yqw, Talbot:2019okv}.
In Fig.~\ref{fig:original_gap_posterior} the gray curves show the resulting posterior distribution for the hyper-parameters that describe the spin magnitude and tilt distributions.
As alluded to above, we choose to show these hyper-parameters as they best demonstrate how our inferences can be biased due to variance cuts in this example.
This plot also includes the distribution of $\sigma_\mathcal{L}^2$ values associated with each of the hyper-posterior samples.
We see that the distribution rails against the threshold of $\sigma_\mathcal{L}^2 = 1$, indicating that a significant portion of potential hyper-posterior samples are being rejected due to variance cuts.
As we will see later, this is biasing our results.

It is not uncommon to use much less than $10^4$ samples to represent a gravitational-wave posterior.
For example, the recommended PE data products from Refs.~\cite{LIGOScientific:2026ctl, LIGOScientific:2025pvj} use as few as $\approx 2 \times 10^3$ posterior samples for some events.
To demonstrate how problematic this can be, we redo our population inference, this time truncating our dataset to use only $2 \times 10^3$ posterior samples for each event.
The results are over-plotted in Fig.~\ref{fig:original_gap_posterior} using red.
We see a significant difference between the red and gray posteriors and a distribution of $\sigma_\mathcal{L}^2$ peaking sharply at the threshold of $1$, implying that the $2 \times 10^3$ samples are not enough for reliable hierarchical inference.

\section{Short-term solutions}\label{sec:easy_fixes}
In this section we discuss techniques that can be used to minimize numerical error in population analyses with existing tools and minimal development of new resources.

\subsection{Rectangular arrays with varying numbers of samples}
For computationally efficient evaluation of the population likelihood in Eq.~\ref{eq:population_likelihood_mc}, we need to compute the Monte-Carlo integral for each event in parallel.
This means the $n$ posterior samples for each event need to be combined into a single rectangular array.
Currently, hierarchical inference packages like \textsc{GWPopulation} \citep{Talbot:2024yqw, Talbot:2019okv} combine posterior samples into a single rectangular array by enforcing that $n$ is identical for all $N$ events.
This is achieved by throwing away fiducial samples so that $n$ is determined by the event with the fewest samples available.
However, the number of samples produced during PE can vary drastically between events.
In our example, we lose a median of $\approx 3 \times10^3$ samples for each event if this truncation is performed.

However, there is a simple solution that allows us to employ computationally efficient rectangular arrays without throwing away samples;
we pad the array with \texttt{NaN} values for events with $n$ less than the maximum. 
When the weight function (Eq.~\ref{eq:population_weights}) is passed \texttt{NaN} values, it returns zero, and so these pseudo-samples are effectively ignored when calculating the likelihood in Eq.~\ref{eq:population_likelihood_mc} (taking care to ensure $n$ is the number of real samples for each event when taking the expectation value).
We have implemented minor adjustments to the \textsc{GWPopulation} and associated \textsc{GWPopulation\_pipe} codebases to do this, which we provide at Refs.~\cite{gwpopulation_padded, gwpopulation_pipe_padded}.

We also note a different method for including unequal numbers of samples for each event implemented in the \textsc{gwax} codebase \citep{gwax}, which further optimizes for computational efficiency.
Namely, to avoid storing and running calculations on placeholder \texttt{NaN} values, \textsc{gwax} combines the posterior samples for all events into a single one-dimensional array.
This one-dimensional array can then be manipulated by keeping track of where each events' samples begin and end, and slicing the array in accordance.

\begin{figure*}    \includegraphics[width=0.85\textwidth]{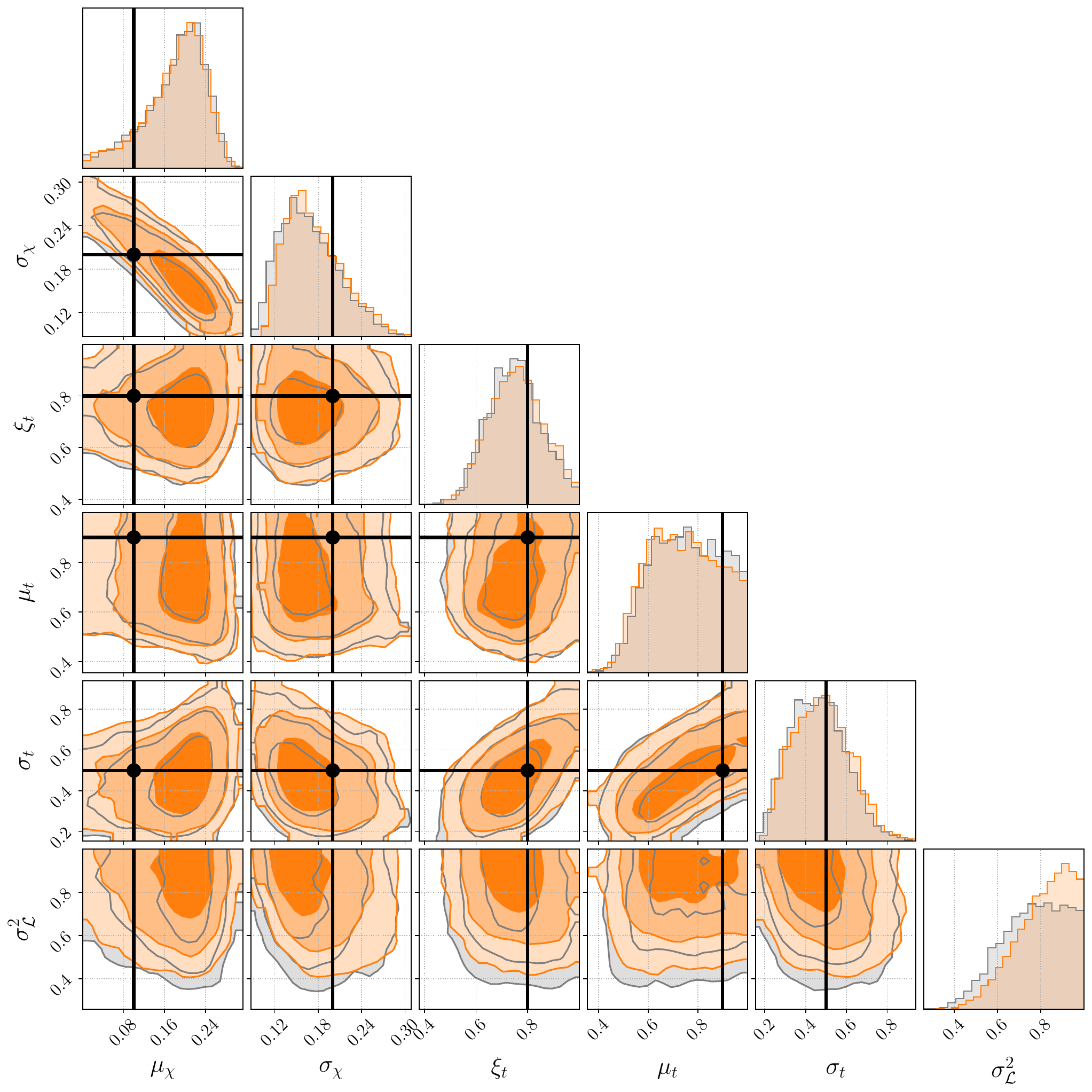}
    \caption{
    Posteriors for spin distribution hyper-parameters, along with the distribution of likelihood variances $\sigma_\mathcal{L}^2$ in the hyper-posterior samples.
    Orange is calculated using a truncated set of $\approx 1\times10^4$ PE samples for each event.
    Meanwhile, gray is calculated using the maximum number of available fiducial samples for each event (ranging from $\approx 1\times10^4{-}2 \times 10^4$).
    From lightest to darkest, the shades in two-dimensional panels indicate 99\%, 90\%, and 50\% credible regions.
    The true injected value is shown in black.
    By using all the available samples, the distribution shifts slightly, most notably in the $(\mu_t,\sigma_t)$ plane.
    The distribution of likelihood variances also peaks less sharply around the threshold of $\sigma_\mathcal{L}^2=1$.
    }
    \label{fig:untruncated_gap_posterior}
\end{figure*}

In Fig.~\ref{fig:untruncated_gap_posterior} we plot the hyper-posteriors (and likelihood variances) calculated two ways: orange uses the truncated set of $\approx 10^4$ fiducial samples for each event while gray uses the full number available.
By using all the available samples, the hyper-posterior changes slightly, and the distribution of likelihood variances peaks less sharply at the threshold.

\subsection{Using nested samples}
\begin{figure}
    \centering
    \includegraphics[width=0.8\columnwidth]{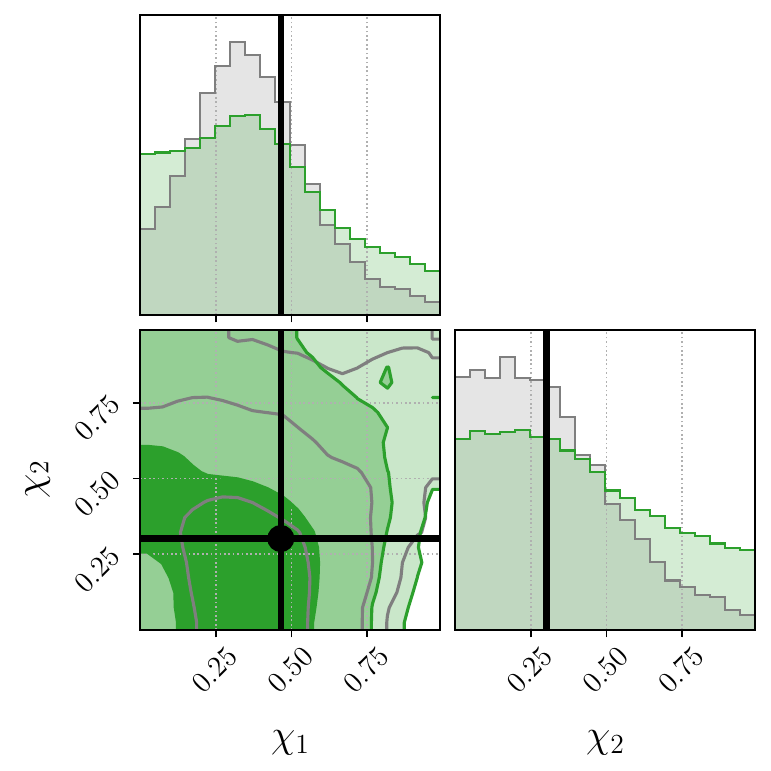}
    \caption{
    Comparison between the distributions of posterior samples (gray) and nested samples (green) for dimensionless spin magnitudes $(\chi_1,\chi_2)$ obtained during PE for a random event from our mock catalog.
    From lightest to darkest, the shades in the two-dimensional panel indicate 99\%, 90\%, and 50\% credible regions.
    The true injected value for the given event's spins is shown in black.
    Notice the distribution of nested samples peaks in the same location as the posterior, but has thicker tails.
    }
    \label{fig:posterior_vs_nested_event}
\end{figure}

Parameter estimation is often performed using a nested sampler like \textsc{Dynesty} \citep{Speagle:2019ivv}.
Nested samplers do not solely produce posterior samples; rather, they sample $\theta$ with the aim of estimating the Bayesian evidence $\mathcal{Z}_{\O}(d_i)$.
The resulting distribution of these ``nested samples'' resembles the posterior, but with heavier tails \citep{Skilling:2006gxv}.
In Fig.~\ref{fig:posterior_vs_nested_event}, we plot an example of this using the spin magnitudes $(\chi_1,\chi_2)$ measured for a random event in our mock catalog.
These nested samples are subsequently rejection sampled (using weights produced during nested sampling) to obtain a batch of unweighted posterior draws \citep{Ashton:2018jfp, Romero-Shaw:2020owr}.
In the case of our mock population, this rejection sampling results in $\approx 60{-}90\%$ of the nested samples being discarded for any event.

While the nested samples that are discarded are typically less useful than the retained ones, they still contain valuable information about the posterior, which we can use to improve our representation of the likelihood surface.
From our mock population, the median number of samples retained after rejection sampling is $\approx 1.3 \times 10^4$.
Meanwhile, the median effective sample size of the weighted nested samples is $\approx 2 \times 10^4$.

Here, we explain how to carry out a population analysis using the nested samples produced during fiducial PE.
Assuming $\theta_i^k$ is now a set of $\hat{n}$ nested samples, we make a simple adjustment to the weights defined in Eq.~\ref{eq:population_weights}:
\begin{equation}
    \hat{w}(\theta_i^k|\Lambda) = W_i^k \, 
    w(\theta_i^k|\Lambda).
\end{equation}
Here, we have multiplied the weight for each sample by a factor $W_i^k$, which is the nested sample weight already calculated by samplers like \textsc{Dynesty} \citep{Speagle:2019ivv}.
Formally, $W_i^k$ is a ratio between the fiducial posterior distribution and the distribution of nested samples:
\begin{align}
    W_i^k = \frac{p_{\O}(\theta_i^k|d_i)}{p_\text{nest}(\theta_i^k|d_i)} .
\end{align}
We then replace $w(\theta_i^k|\Lambda)$ with $\hat{w}(\theta_i^k|\Lambda)$ and $n$ with $\hat{n}$ in Eq.~\ref{eq:population_likelihood_mc} to calculate the population likelihood using the entirety of the nested samples obtained during PE.
Our adjustments to \textsc{GWPopulation\_pipe} \citep{gwpopulation_pipe_padded} add the option to use weighted fiducial samples, enabling population inference with nested samples as described above.

\begin{figure*}
    \includegraphics[width=0.85\textwidth]{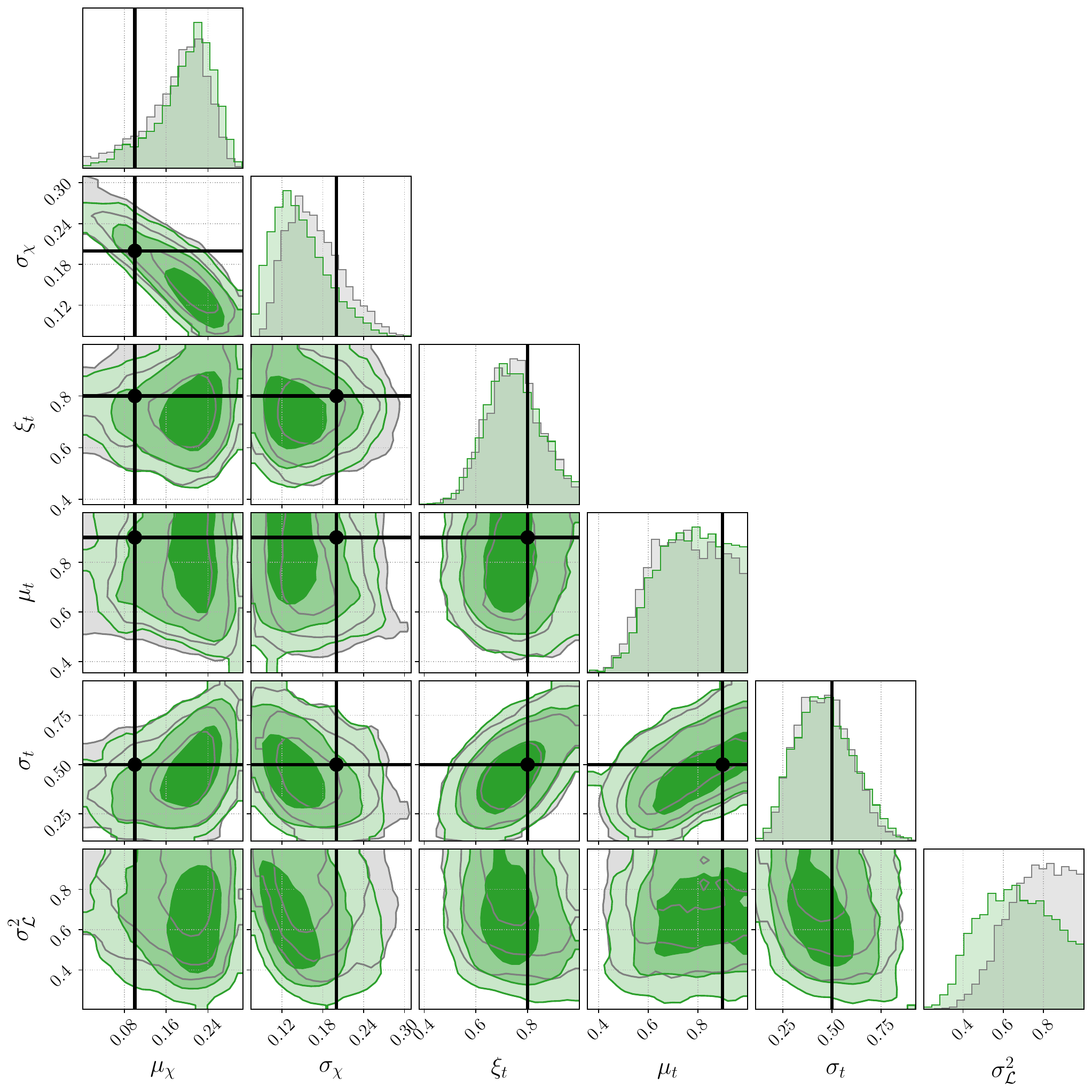}
    \caption{
    Posteriors for spin distribution hyper-parameters, along with the distribution of variances in the population likelihood $\sigma_\mathcal{L}^2$.
    Gray shows the hyper-posterior inferred using $\approx 1 \times 10^4{-}2 \times 10^4$ unweighted posterior samples for each event.
    Meanwhile, green shows the hyper-posterior inferred using the entire batch of weighted nested samples (that were rejection sampled to produce the unweighted posterior samples).
    From lightest to darkest, the shades in two-dimensional panels indicate 99\%, 90\%, and 50\% credible regions.
    The true injected value is shown in black.
    We see the green contours inferred using nested samples shift compared to our benchmark results in gray, particularly in $\sigma_\chi$.
    When using nested samples, the distribution of likelihood variances no longer peaks at the threshold of $\sigma_\mathcal{L}^2$, although it is clear that a significant number of hyper-posterior samples are still being rejected due to reweighting uncertainties.
    }
    \label{fig:nested_gap_posterior}
\end{figure*}

In Fig.~\ref{fig:nested_gap_posterior}, we show the posterior for the spin hyper-parameters (and the distribution of likelihood variances) calculated two ways.
The result in gray uses our benchmark $\approx 1 \times 10^4{-}2 \times 10^4$ posterior samples for each event.
The result in green uses importance sampling with the $\approx 3\times10^4{-}1\times10^5$ nested samples that produced the aforementioned posterior draws via rejection sampling.
The differences in the posteriors are more significant than those in Fig.~\ref{fig:untruncated_gap_posterior}, and the distribution of $\sigma_\mathcal{L}^2$ values no longer peaks at the threshold of $1$ when using the nested samples.
Thus, nested samples provide a method to further reduce numerical error in population analyses.

One might wonder if the broadness of the distribution of nested samples can itself provide an advantage in population reweighting.
Naively, we might expect that when some events are reweighted to follow the population prior $\pi(\theta|\Lambda)$, it is possible that the bulk of the events' posterior support shifts to regions of parameter space in which the fiducial posteriors $p_{\O}(\theta|d_i)$ have relatively little support.
In this case, our samples being distributed more broadly than the fiducial posterior may be advantageous---resulting in a smaller value for the bracketed term in Eq.~\ref{eq:population_uncertainty} than an equivalent number of posterior samples.
We test this by reweighting each event in our mock catalog to follow the true population model, once using nested samples, and a second time using an equivalent number of posterior samples, then calculating the resulting effective sample size and reweighting variance for each.
From our 300 mock events, we do not find any instances in which the nested samples outperform the posterior samples.
This indicates, at least for this specific problem, that samples distributed according to the fiducial posterior $p_{\O}(\theta|d_i)$ are more effective for population inference than those following the distribution of nested samples $p_\mathrm{nest}(\theta|d_i)$.

\section{Long-term solutions with normalizing flows}\label{sec:flows}
We now propose a solution for instances in which the methods outlined in Section~\ref{sec:easy_fixes} fail to reduce the variance sufficiently.
Once we have squeezed as much information as possible out of the nested samples, the only path forward is to generate additional posterior samples.
In Section~\ref{sec:discussion}, we discuss the possibility of replacing importance sampling entirely with neural networks, but even then, we believe large batches of posterior samples will be needed to train these approximations.
Thus, to move forward, we seek an efficient method to generate additional posterior samples.

One might be tempted to create additional samples by running more fiducial parameter estimation.
However, naive generation of fiducial samples is inefficient because the sampler draws proposal samples from a broad, uninformative prior.
Furthermore, traditional sampling is not highly parallelizable.
Depending on signal duration and SNR, generating the initial $\approx 1 \times 10^4 - 2 \times 10^4$ posterior samples took running three instances of the sampler in parallel, each utilizing 8 CPU cores, between $\approx 1$ hour to several days of run time.

However, we can take advantage of the fact that we already have ${\cal O}(10^4)$ posterior samples (or up to ${\cal O}(10^5)$ nested samples with weights) pointing us to the bulk of the posterior support.
In other words, we do not need to do inference from scratch given that we already have information about the shape of the posterior.
To that end, we propose using normalizing flows \citep{Papamakarios:2019fms} trained on existing posterior (or nested) samples to create a neural posterior estimator (NPE) for each event.
Using this NPE, millions of samples approximating the posterior can be drawn within seconds, which can be re-weighted in parallel to follow the exact posterior.
These weighted draws can be added back into the training set to iteratively improve the accuracy of the NPE.

This is similar to pipelines like \textsc{Dingo} \citep{Green:2020dnx, Dax:2021tsq, Dax:2022pxd}, which aim to replace traditional gravitational-wave PE by training normalizing flows that are conditional on strain data $d$.
The difference is that \textsc{Dingo} aims to obtain a flow representation of the posterior \textit{using the raw strain data} whereas we focus on the simpler problem of obtaining a flow representation based on existing posterior samples---however they may have been obtained.
We train unconditional NPEs for each event on samples from PE as we find this technique conceptually simpler and easier to experiment with, while also producing more accurate NPEs.
We discuss this further in Section~\ref{sec:discussion}.
We note that Ref.~\citep{Dax:2022pxd} also employs unconditional normalizing flows fit to existing samples as part of its importance sampling framework.
Specifically, an unconditional flow is fit to proxy parameters (used in their group-equivariant NPE algorithm) in order to calculate the proposal density when importance sampling draws of $\theta$ from their conditional, simulation-based flow.
A closer analogue to our approach is also demonstrated in Ref.~\citep{Dax:2022pxd}, where the authors train an unconditional NPE for GW151012, using $\approx 10^6$ posterior samples generated via \textsc{Bilby} and \textsc{Dynesty}, against which they benchmark the performance of their simulation-based approach.
While sharing the same core mechanism, as will be outlined in the remainder of this section, our methodology differs in two ways: our flows are trained using nested samples to inform areas of low density without the need for excessive numbers of training samples, and our flows are iteratively trained on their own outputs to refine their accuracy.
These variations reflect a difference in the intended applications of unconditional NPEs in these two works.
Whereas Ref.~\citep{Dax:2022pxd} uses these to validate their simulation-based approach, we aim to use them as a practical way of efficiently generating large batches of samples for use in hierarchical inference.

Our proposal also bears similarities to the recently introduced \textsc{ASPIRE} pipeline \citep{Williams:2025aar} and the work of Ref.~\citep{Prathaban:2026kft} which, like our method, revolve around training normalizing flow models on existing gravitational-wave posteriors.
Whereas we use our NPEs to generate samples for use in hierarchical inference (measuring population properties), \textsc{ASPIRE} uses NPEs to accelerate sequential inference (re-measuring event-level properties under new assumptions).
For example, using an alternative waveform model or introducing additional parameters like precession or eccentricity.
Similarly, Ref.~\citep{Prathaban:2026kft} uses NPEs trained on approximate PE results from the \texttt{simple-pe} algorithm \citep{Fairhurst:2023idl} to accelerate full PE using nested sampling.
Also see Refs.~\citep{Wouters:2025zju, Golomb:2021tll, DEmilio:2021laf, Wysocki:2020myz} for other works that utilize density estimates for gravitational-wave posteriors.

We provide code for building, training and sampling normalizing flows as described throughout the remainder of this section in Ref.~\cite{flow_repo}.

\subsection{Initial flow construction}
\begin{figure*}
    \begin{tikzpicture}[
  scale=0.9, transform shape,
  flowbox/.style={
    draw=red!80!black, thick, rectangle,
    minimum width=3.2cm, minimum height=0.75cm,
    align=center, font=\small
  },
  transformbox/.style={
    draw=green!80!black, thick, rectangle,
    minimum width=3.2cm, minimum height=0.75cm,
    align=center, font=\small
  },
  endbox/.style={
    draw=blue!70!black, thick, rectangle,
    minimum width=3.2cm, minimum height=0.75cm,
    align=center, font=\small
  },
  mlpbox/.style={
    draw=red!80!black, thick, rectangle,
    minimum width=2.6cm, minimum height=1.6cm,
    align=center, font=\small
  },
  arr/.style={-{Stealth[scale=1.1]}, thick},
  lbl/.style={font=\scriptsize, fill=white, inner sep=1.5pt}
]

    \node[endbox]  (base)    {$\mathcal{N}(u^0| 0,1)$};
    \node[font=\scriptsize, black, above=1pt of base] (baselab) {Base distribution};

    \node[flowbox, below=1.0cm of base](split){Coupling Split};
    \node[flowbox, below=0.8cm of split](tanh){Tanh};
    \node[flowbox, below=0.8cm of tanh](spline)
    {Spline$(v^j_\mathrm{t}|x^j_\mathrm{c})$};
    \node[flowbox, right=2.0cm of spline](mlp){MLP$(u^j_\mathrm{c}|\lambda^j_{\mathrm{MLP},i})$\\[2pt]Width $= 128$\\Depth $= 8$\\Activation $=$ SiLU};
    \node[flowbox, below=0.8cm of spline](tanhInv){Tanh$^{-1}$};
    \node[flowbox, below=0.8cm of tanhInv](join){Coupling Join};
    \node[flowbox, below=0.8cm of join](permute){Permute};
    
    \node[transformbox, below=1.0cm of permute](affine){Affine$(u^{16}|\lambda_{\mathrm{aff},i})$};
    \node[transformbox, below=0.8cm of affine](sinhArcsinh){SinhArcsinh$({u^{16}}^*|\lambda_{\mathrm{SAS},i})$};
    \node[transformbox, below=0.8cm of sinhArcsinh](sigmoid){Sigmoid};
    \node[transformbox, below=0.8cm of sigmoid](moveBound){Affine$({u^{16}}^{***}|\theta_{\min}, \theta_{\max}-\theta_{\min})$};

    \node[endbox, below=1.0cm of moveBound]  (target)    {$q(\theta|\lambda_i)$};
    \node[font=\scriptsize, black, below=1pt of target] (targetlab) {Approximate posterior distribution};

    \draw[arr](base) -- node[lbl, right]{$u^0$} (split);
    \draw[arr](split) -- node[lbl, right]{$u^j_\mathrm{t}$} (tanh);
    \draw[arr](split.east) -| node[lbl, pos=0.25, below]{$u^j_\mathrm{c}$} (mlp.north);
    \draw[arr](tanh) -- node[lbl, right]{$v^j_\mathrm{t}$} (spline);
    \draw[arr](mlp.west) -- node[lbl, below]{$x^j_\mathrm{c}$} (spline.east);
    \draw[arr](spline) -- node[lbl, right]{$\bar{v}^j_\mathrm{t}$} (tanhInv);
    \draw[arr](tanhInv) -- node[lbl, right]{$\bar{u}^j_\mathrm{t}$} (join);
    \draw[arr](split.east) -- ++(5.4cm,0) |- (join.east);
    \draw[arr](join) -- node[lbl, right]{$\bar{u}^j$} (permute);
    \draw[arr](permute.west) -- ++(-1cm,0) |- node[lbl, pos=0.25, left]{$u^{j+1}$} (split.west);
    \draw[arr](permute) -- node[lbl, right]{$u^{16}$} (affine);
    \draw[arr](affine) -- node[lbl, right]{${u^{16}}^*$} (sinhArcsinh);
    \draw[arr](sinhArcsinh) -- node[lbl, right]{${u^{16}}^{**}$} (sigmoid);
    \draw[arr](sigmoid) -- node[lbl, right]{${u^{16}}^{***}$} (moveBound);
    \draw[arr](moveBound) -- node[lbl, right]{$\theta$} (target);

    \begin{scope}[on background layer]
      \node[
        draw=red, thick,
        rounded corners=7pt,
        fit=(split)(tanh)(spline)(mlp)(tanhInv)(join)(permute),
        inner sep=13pt
      ] (flowlayerbox) {};
    \end{scope}
    \node[font=\scriptsize, black, below=1pt of flowlayerbox.south]
      {Flow Layer $j \in \{0,...,16\}$};
    
\end{tikzpicture}
    \caption{
    Diagram of our normalizing flow architecture.
    We start with a zero-mean, unit-variance normal base distribution.
    The normally distributed latent variable goes through 16 layers of neural spline flow transformations.
    In each, the 13-dimensional latent variable $u^j$ is split into 6 conditioning $u^j_c$ and 7 transformed $u^j_t$ variables.
    The conditioning variables act as inputs for a neural network (multi-layer perceptron) of width 128 and depth 8, using SiLU activation functions.
    The outputs of this network are the heights of the spline knots (bounded from -1 to 1), used to shift the transformed variables.
    Meanwhile, the transformed variables are passed through a tanh function to rebound them between $(-1,1)$ where the spline knots are located.
    After the spline transformation, an inverse tanh function ``undoes'' the $(-1,1)$ bounding.
    Finally, these transformed variables are recombined with the (unchanged) conditioning variables, and their order is shuffled by a random permutation (fixed at initialization).
    After 16 flow layers, the resulting latent variables go through a series of ``post-processing'' transformations.
    Firstly, an affine transformation undoes any regularization.
    Then, a SinhArcsinh function allows for stretching of the tails.
    A sigmoid transformation then transforms the unbounded latent space into one bounded from $(0,1)$, after which, a final fixed affine transformation shifts these bounds to line up with the prior boundaries implied by $\pi_{\O}(\theta)$.
    }
    \label{fig:flow_architecture}
\end{figure*}
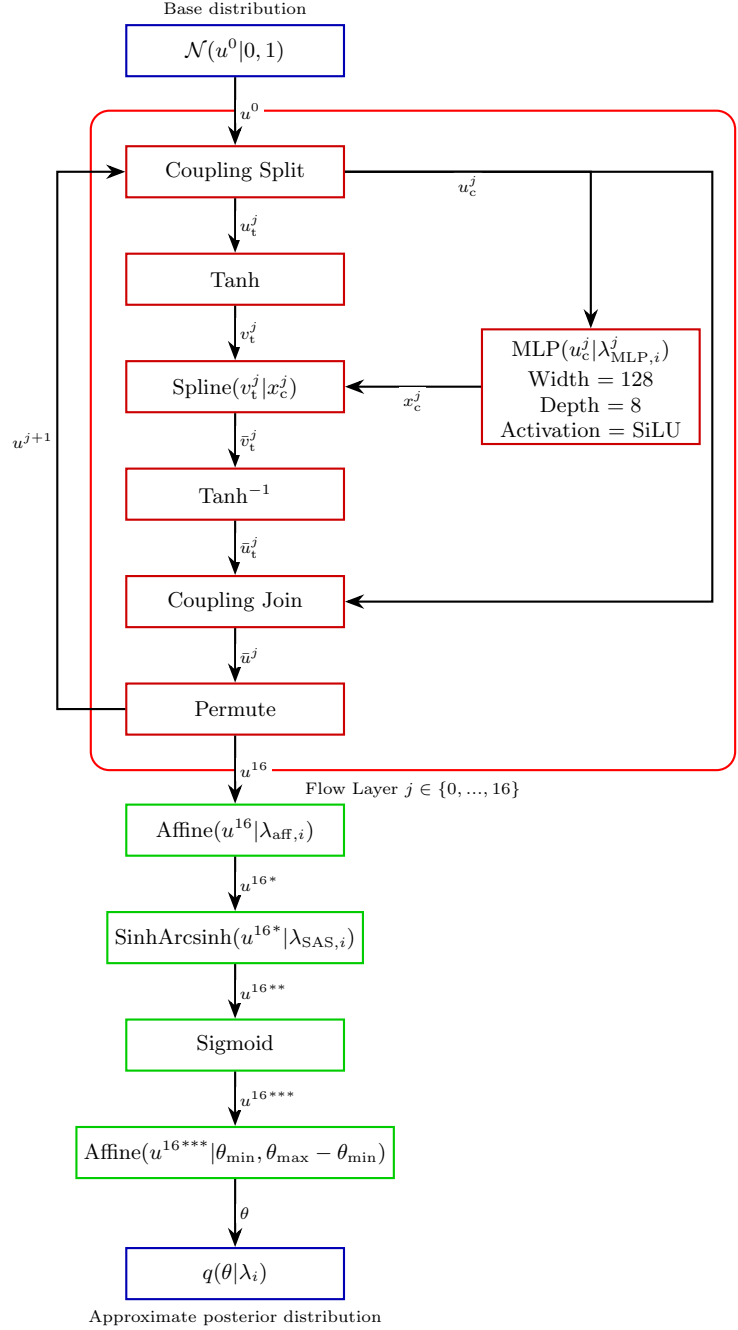

We use the \textsc{FlowJax} package \cite{ward2023flowjax} to construct, fit and sample from our NPEs.
Within \textsc{FlowJax}, we construct a 13-dimensional (marginalizing phase, time and distance, but including time-jitter; see Ref.~\cite{Romero-Shaw:2020owr}) neural spline flow \citep{Durkan:2019nsq} with 16 layers.
We illustrate the specifics of the architecture in Fig.~\ref{fig:flow_architecture}.
We also experimented with Student's-$t$ and multi-modal base distributions, as well as enforcing periodic boundary conditions in appropriate parameters \citep{Rezende:2020hrd}, but we did not see any noteworthy improvement in performance with these modifications.
Given the general structure provided in Fig.~\ref{fig:flow_architecture}, we also experimented with different neural network depths and widths, as well as different numbers of flow layers, finding those shown in the diagram to be the best performing when averaging across the entire mock catalog.

We initially train the NPEs using the nested samples obtained during fiducial posterior sampling.
Using the entire set of weighted nested samples rather than just rejection-sampled posterior draws has the added advantage of better informing the tails of the NPE (see also Ref.~\cite{Prathaban:2024rmu}).
We use the Kullback-Leibler (KL) divergence as a loss function:
\begin{equation} \label{eq:loss}
    D_\mathrm{KL}\left(p_{\O}(\theta|d_i) || q(\theta|\lambda_i)\right) = \int d\theta \ p_{\O}(\theta|d_i) \ln \left[\frac{p_{\O}(\theta|d_i)}{q(\theta|\lambda_i)}\right],
\end{equation}
where $q(\theta|\lambda_i)$ is the NPE probability, and $\lambda_i$ are the hyper-parameters that govern its shape for a given event.
We perform a Monte-Carlo integral to approximate Eq.~\ref{eq:loss} using $\hat{n}$ nested samples:
\begin{equation}\label{eq:loss_mc}
    D_\mathrm{KL} \approx -E_{\hat{n}} \left[ W_i^k \ln q(\theta_i^k|\lambda_i)\right] \times C.
\end{equation}
Here, $C$ is some factor that remains constant for any value of $\lambda_i$, so can safely be ignored during training.
Note we have dropped the parentheses after $D_\mathrm{KL}$ for brevity.
We set aside $10\%$ of our data for validation.
We use a learning rate of $10^{-4}$, with a batch size of $128$ training samples.
Optimization is performed with an \texttt{adamW} optimizer, clipping any batch losses that are a factor of $\geq 5$ times the global norm for stability.
Training is terminated when the validation loss has not improved for $10$ steps in a row.
Training takes $\approx 30$ minutes for a typical event on a single CPU core or $\approx 5$ minutes on a GPU (benchmarked using an \texttt{NVIDIA GeForce RTX 4070 Ti}).

\subsection{Flow sampling and validation}
Once an NPE has been trained, we generate $\eta = 10^6$ samples of $\theta$, also computing the draw probability $q(\theta|\lambda_i)$ of each.
This takes $\mathcal{O}(10)$ seconds.
For each sample, we compute the target likelihood $\mathcal{L}(d_i|\theta)$ and (fiducial) prior probability $\pi_{\O}(\theta)$ using \textsc{Bilby}.
This allows us to calculate a weight for each sample:
\begin{equation}\label{eq:flow_weight}
    w_{\mathrm{NPE},i} \propto \frac{p_{\O}(\theta_i|d_i)}{q(\theta_i|\lambda_i)}.
\end{equation}
Calculating these likelihoods and weights for all $\eta = 10^6$ samples takes $\approx 5-30$ minutes using 16 CPU cores, depending primarily on the event's signal duration.

\begin{figure*}
    \includegraphics[width=\textwidth]{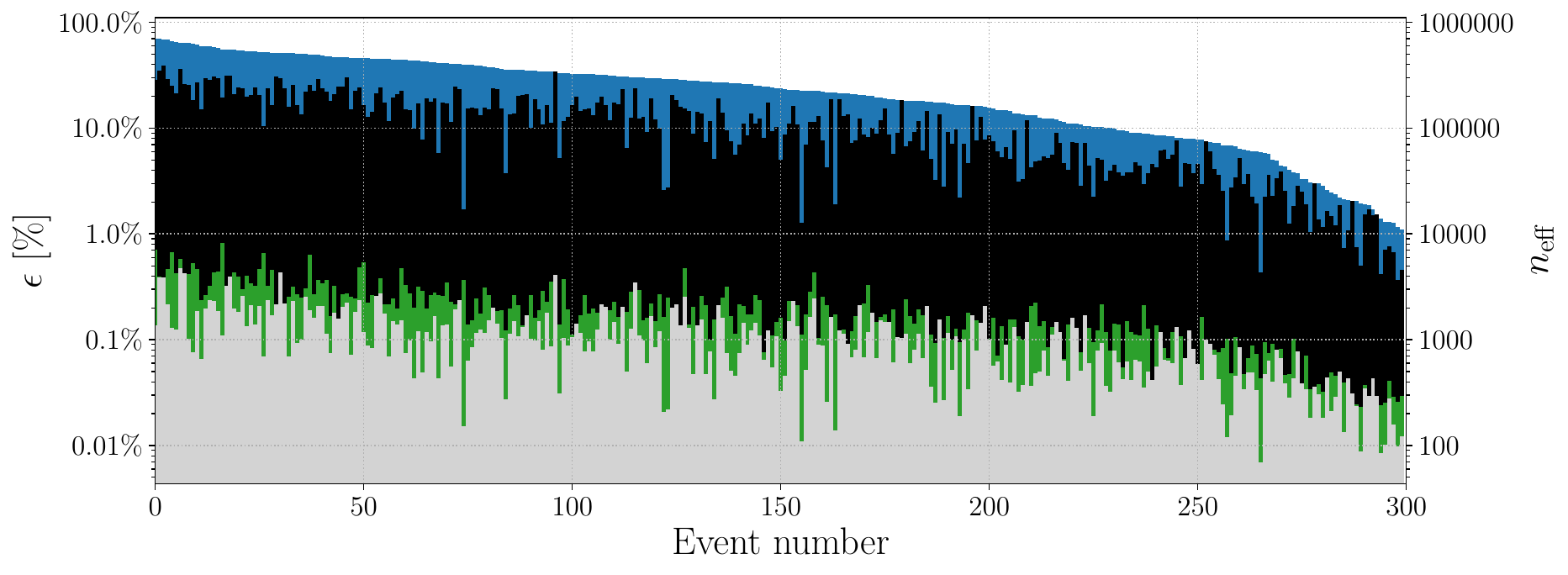}
    \caption{
    Bar chart of the normalizing flow performance for each of the 300 mock events.
    The height of each blue and black bar indicates the effective sample size $n_\mathrm{eff}$ of a batch of $\eta = 10^6$ NPE samples, that have been reweighted to follow the fiducial posterior $p_{\O}(\theta|d_i)$ for the given event.
    This value can also be represented as an efficiency $\epsilon = n_\mathrm{eff}/\eta$, as is displayed on the left axis.
    For each event, the black and blue bars represent the efficiency before and after iterative training respectively.
    Thus, the visible portion of the blue bars can be thought of as the improvements in NPEs' performances as a result of iterative training.
    Events are numbered and sorted according to the effective sample size achieved by the associated NPE after iterative training (best to worst).
    Meanwhile, the gray and green bars show the number of samples left after rejection sampling draws from the initial NPE and the iteratively trained NPE respectively.
    }
    \label{fig:flow_metrics}
\end{figure*}

Using these weights, we calculate the effective sample size 
\begin{equation}
    n_\mathrm{eff} = \frac{\left(\sum_k^\eta  w_{\mathrm{NPE},i}^k\right)^2}{\sum_k^\eta  \left(w_{\mathrm{NPE},i}^k\right)^2} ,
\end{equation}
and the associated efficiency $\epsilon = n_\mathrm{eff}/\eta$.
The efficiency acts as our primary diagnostic tool in assessing the quality of a given NPE.
We plot this initial efficiency across events in blue in Fig.~\ref{fig:flow_metrics}, finding a median efficiency of 10\%, a minimum of 0.4\%, and a maximum of 39\%.

\subsection{Iterative training}

The accuracy of an NPE (especially in high-dimensional cases like these), is often limited by the size of the training dataset \cite{Frazier:2024}.
As such, we can improve the fit of our NPEs by taking the newly generated samples, mixing them into our training set, and repeating the process.

To do so, we need to define a new loss function.
In Eq.~\ref{eq:loss_mc}, we take samples from a nested proposal distribution, then reweight them to follow the posterior target distribution in order to calculate a KL divergence via Monte-Carlo integration.
Meanwhile, our new NPE samples come from a different proposal distribution defined by our normalizing flow.
In theory, we could adjust the loss function so that the proposal distribution (the denominator in the weights) now represents a mixture model between the nested distribution and the flow distribution.
While we can easily compute the flow probability (thus the aforementioned mixed weights) for the nested samples, we do not have a method to compute the nested draw probability for the new NPE samples.
Thus, this method is untenable.

We instead compute the Monte-Carlo integral in Eq.~\ref{eq:loss_mc} separately with the nested samples, then again with the new NPE samples.
This gives two estimates for the integral in Eq.~\ref{eq:loss}:
\begin{align}
    D_\mathrm{KL,nest} &\approx E_{\hat{n}} \left[ W_i^k \ln \left[\frac{p_{\O}(\theta_i^k|d_i)}{q(\theta_i^k|\lambda_i)}\right]\right], \\
    D_\mathrm{KL,NPE} &\approx E_\eta \left[ w_{\mathrm{NPE},i}^j \ln \left[\frac{p_{\O}(\theta_i^j|d_i)}{q(\theta_i^j|\lambda_i)}\right]\right].
\end{align}
We have used different superscripts $k$ and $j$ to emphasize that these are calculated on two different sets of samples.
Also note that we no longer factor out the posterior as a constant, as it can no longer be treated as such in the next step.

We can then combine these two estimates into a single loss value by taking their average, each weighted by the inverse variance in the estimator:
\begin{equation}\label{eq:iterative_loss}
     D_\mathrm{KL} \approx \frac{\sigma_\mathrm{nest}^{-2}}{\sigma_\mathrm{nest}^{-2} + \sigma_\mathrm{NPE}^{-2}} D_\mathrm{KL,nest} + \frac{\sigma_\mathrm{NPE}^{-2}}{\sigma_\mathrm{nest}^{-2} + \sigma_\mathrm{NPE}^{-2}} D_\mathrm{KL,NPE},
\end{equation}
where
\begin{multline}
    \sigma_\mathrm{nest}^{2} = \frac{1}{\hat{n}} \Bigg( E_{\hat{n}} \left[ \left(W_i^k \ln \left[\frac{p_{\O}(\theta_i^k|d_i)}{q(\theta_i^k|\lambda_i)}\right]\right)^2\right] - \\ E_{\hat{n}} \left[ W_i^k \ln \left[\frac{p_{\O}(\theta_i^k|d_i)}{q(\theta_i^k|\lambda_i)}\right]\right]^2 \Bigg),
\end{multline}
and
\begin{multline}
    \sigma_\mathrm{NPE}^{2} = \frac{1}{\eta} \Bigg( E_{\eta} \left[ \left(w_{\mathrm{NPE},i}^j \ln \left[\frac{p_{\O}(\theta_i^j|d_i)}{q(\theta_i^j|\lambda_i)}\right]\right)^2\right] - \\ E_{\eta} \left[ w_{\mathrm{NPE},i}^j \ln \left[\frac{p_{\O}(\theta_i^j|d_i)}{q(\theta_i^j|\lambda_i)}\right]\right]^2 \Bigg).
\end{multline}
As more iterations are performed, the training set can be further supplemented, and the loss function in Eq.~\ref{eq:iterative_loss} can be extended to include additional terms for each batch of NPE samples.

For all events in the mock catalog, we perform two additional iterations on the normalizing flow training using the above method, such that the final flow is trained with $\hat{n}$ nested samples and $2\eta$ flow samples drawn from two different NPEs.
To speed up training and ensure stability, we increase the training batch size to 2048 for these iterations.
For the 3 events that are left with NPE sampling efficiencies $<1\%$, we continue iteratively training until this threshold is reached.
This requires 1-2 additional iterations.

Due to stochasticity and occasional instabilities during training, the most efficient iteration of an event's NPE is not always the final one.
We instead proceed with whichever iteration of an event's NPE produces the highest efficiency batch of samples.
The result is a median sampling efficiency of 24\% across the catalog, with a minimum of 1\%, and a maximum of 70\%.
The efficiency for each event is plotted in Fig.~\ref{fig:flow_metrics} in blue.

\subsection{Obtaining unweighted flow samples}\label{sec:unweighted_samples}
In theory, samples from the NPEs (along with their weights) could be used directly in population inference in the same way we used weighted nested samples in Section~\ref{sec:easy_fixes}.
However, even with our relatively high importance sampling efficiencies of $\approx20\%$ across the catalog, this would require representing each event with $>10^5$ NPE samples just to catch up to the accuracy achieved in Section~\ref{sec:easy_fixes}.
From our testing, going much beyond this number of samples becomes computationally prohibitive.

Instead, to optimize computational efficiency, we would like to turn these weighted NPE samples into unweighted batches of posterior samples.
These can then be appended to our existing $\approx 1 \times 10^4{-}2 \times 10^4$ PE samples.
However, in every batch of NPE samples, we find at least one extreme weight $w_{\mathrm{NPE},i}$ that is $\gtrsim 100$ times larger than that of the average sample.
We explore these extreme weights further in Appendix~\ref{sec:diagnostics}.
When the bulk of the distribution is well-modeled, these extreme weights only marginally affect the importance sampling efficiencies reported above.
However, they result in catastrophically inefficient rejection sampling.
In Fig.~\ref{fig:flow_metrics}, gray bars show the number of samples left after rejection sampling the initial batch of NPE samples, and green bars show the number of samples remaining after rejection sampling the final batch of NPE samples (after several iterations of training).
All batches, despite sometimes reaching up to 70\% importance sampling efficiency, lose $>99\%$ of draws after rejection sampling, producing only $\mathcal{O}(100)-\mathcal{O}(1000)$ samples from the $\eta=10^6$ initial draws.

To overcome this, we use a method similar to resample-move steps seen in sequential Monte Carlo algorithms \citep{Gilks:2001, Doucet:2009}.
Specifically, we redraw (with replacement) $1\times10^5$ samples from each batch of $\eta=10^6$ according to their weights.
While the resulting sample sets represent the target posterior distributions, they contain duplicates, and therefore are not independently distributed.
We overcome this by running a short MCMC chain (50 steps) on every sample in order to break the degeneracy between duplicate samples.

There is no rigorous method that we know of to determine if the initially duplicate samples have been sufficiently decoupled.
Naively, we determine this to be when the average distance to a sample's $m$ duplicates has become larger than the average distance to the $m$ nearest non-duplicate samples; indicating that the duplicates have ``mixed in'' with the rest of the draws.
In the few instances where this has not been achieved after 50 steps, we iteratively perform another 50 steps until the threshold has been met.
For a typical event only requiring 50 steps, this takes $\lesssim 30$ minutes with 16 CPU cores.
Meanwhile, some events (particularly those with low efficiencies; $\epsilon \approx 1\%$) may require up to $\approx 2000$ steps, taking $\approx 20$ hours.

\subsection{Supplementing population analyses with flow samples}
\begin{figure*}    \includegraphics[width=0.85\textwidth]{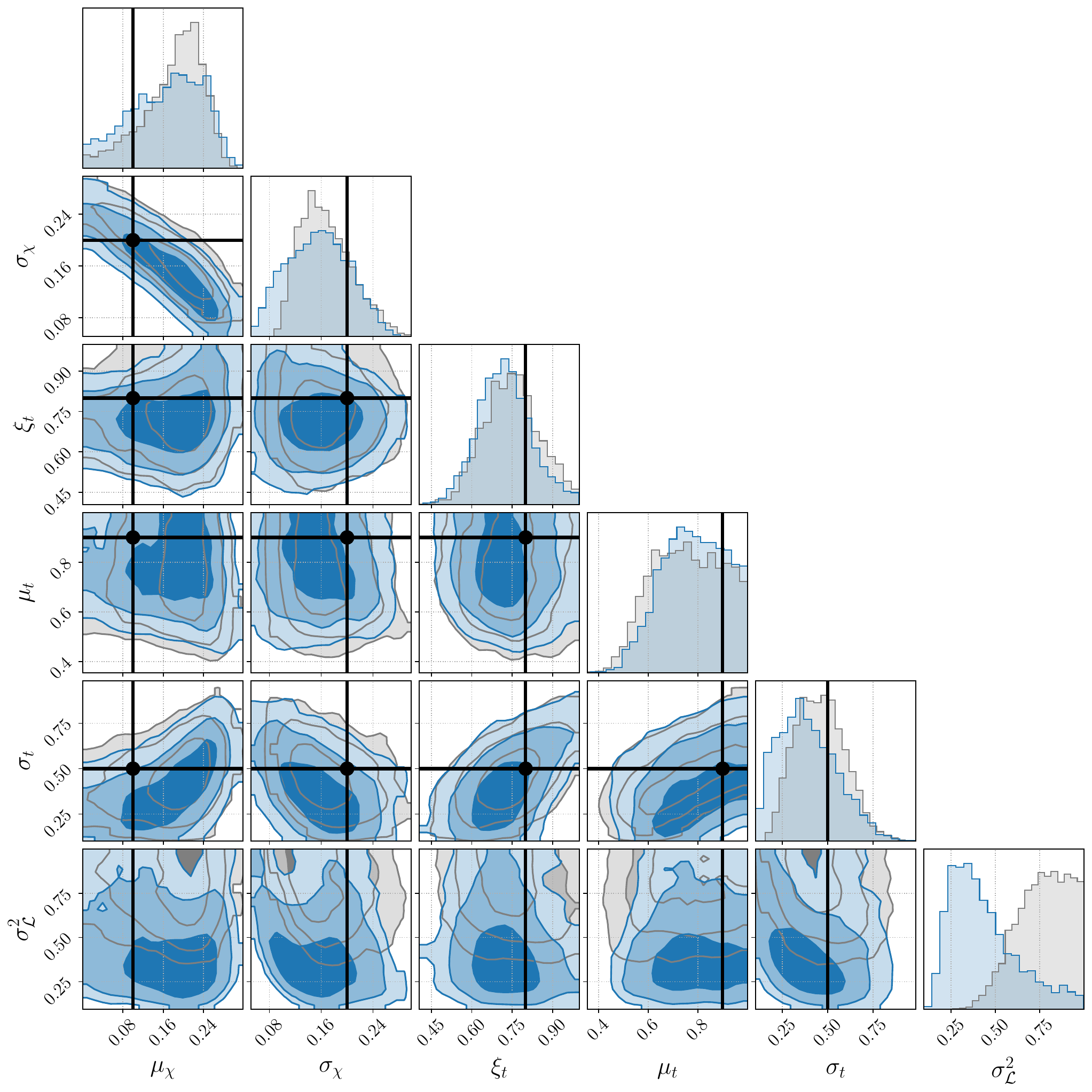}
    \caption{
    Posteriors for spin distribution hyper-parameters, along with the distribution of associated variances in the population likelihood $\sigma_\mathcal{L}^2$.
    Gray shows the hyper-posterior inferred using our original $\approx 1\times 10^4{-}2 \times 10^4$ posterior samples for each event.
    Meanwhile, blue shows the hyper-posterior inferred using sample sets supplemented with new draws from our NPEs so that each event is represented with $5 \times 10^4$ posterior samples.
    From lightest to darkest, the shades in two-dimensional panels indicate 99\%, 90\%, and 50\% credible regions.
    The true injected value is shown in black.
    We see the blue contours differ significantly from the benchmark result in gray.
    The blue distribution of likelihood variances peaks far from the threshold of $\sigma_\mathcal{L}^2 = 1$, only intersecting in the tail.
    This indicates that only a small portion of the hyper-posterior remains inaccessible due to reweighting uncertainties.
    }
    \label{fig:flow_gap_posterior}
\end{figure*}

To illustrate how NPEs can be used to reduce numerical error, we perform population inference two more times.
First, we represent each event with $5 \times 10^4$ posterior samples: supplementing the original $\approx 1\times10^4{-}2\times10^4$ PE samples with $\approx 3\times10^4{-}4\times10^4$ new samples generated with our NPEs.
We plot the resulting hyper-posterior and likelihood variances in Fig.~\ref{fig:flow_gap_posterior}.
We now see an even more significant difference from our benchmark results plotted in gray.
Furthermore, the distribution of likelihood variances now intersects with the threshold of $\sigma_\mathcal{L}^2 = 1$ in the tail, rather than at or near the peak, indicating that relatively few hyper-posterior samples are being rejected due to large reweighting uncertainties.

\begin{figure*}    \includegraphics[width=0.85\textwidth]{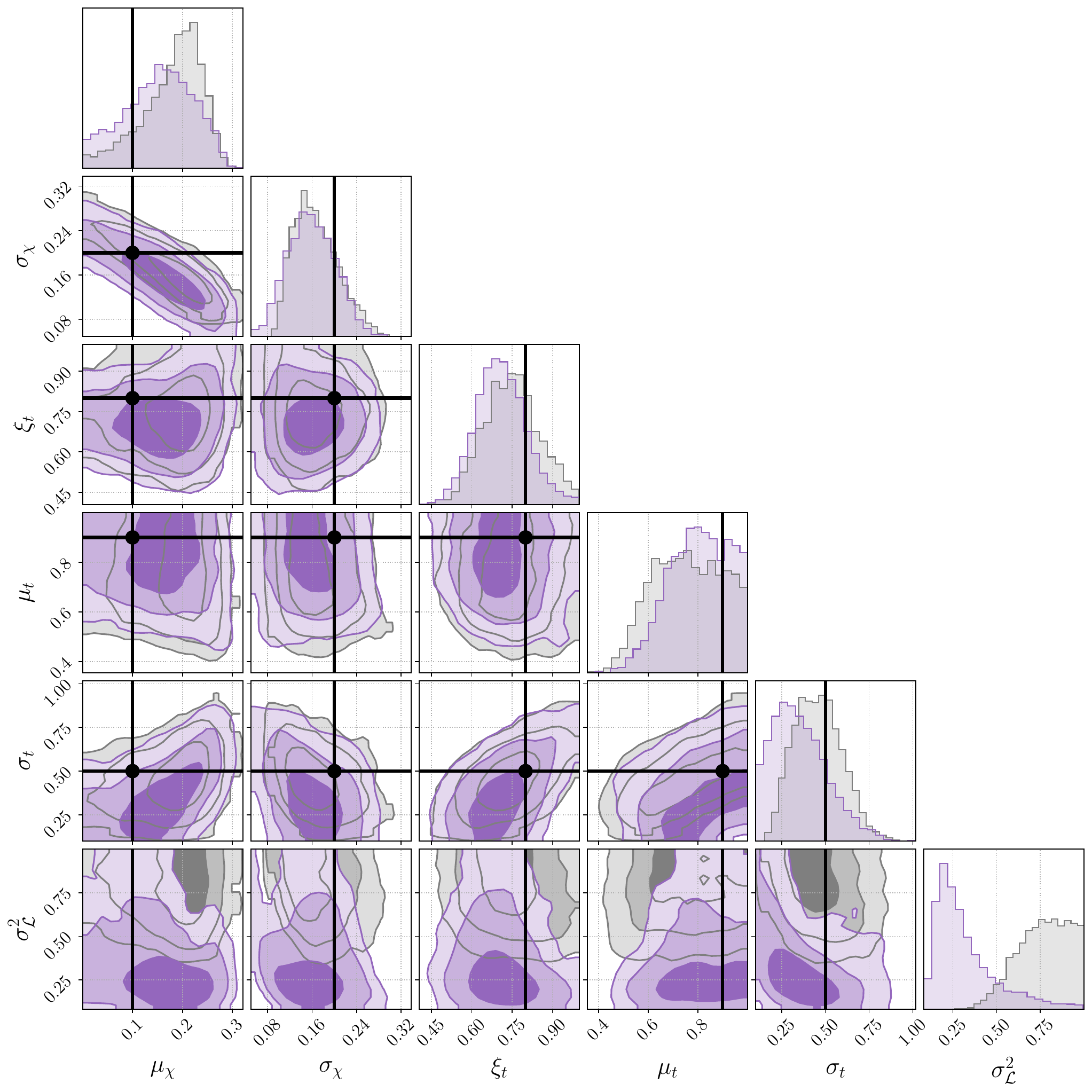}
    \caption{
    Posteriors for spin distribution hyper-parameters, along with the distribution of associated variances in the population likelihood $\sigma_\mathcal{L}^2$.
    Gray shows the hyper-posterior inferred using our original $\approx 1\times 10^4{-}2 \times 10^4$ posterior samples for each event.
    Meanwhile, purple shows the hyper-posterior inferred using sample sets supplemented with new draws from our NPEs so that each event is represented with $1 \times 10^5$ posterior samples.
    From lightest to darkest, the shades in two-dimensional panels indicate 99\%, 90\%, and 50\% credible regions.
    The true injected value is shown in black.
    We see the purple contours look similar to the blue contours from Fig.~\ref{fig:flow_gap_posterior}, inferred using $5 \times 10^4$ samples, with the exception of more support as $\sigma_t$ approaches the lower-bound of the hyper-prior.
    The purple distribution of likelihood variances now has very little overlap with the threshold of $\sigma_\mathcal{L}^2 = 1$, suggesting that reweighting uncertainty is no longer an issue.
    }
    \label{fig:extra_flow_gap_posterior}
\end{figure*}

Finally, we perform population inference representing each event with $1 \times 10^5$ posterior samples: supplementing each event with $\approx 8\times10^4{-}9\times10^4$ new NPE samples.
In Fig.~\ref{fig:extra_flow_gap_posterior}, we show the resulting hyper-posterior in purple along with the distribution of variances in the population likelihood.
We now see very few hyper-posterior samples near the threshold of $\sigma_\mathcal{L}^2 =1$, indicating we can now explore the vast majority of the hyper-posterior without encountering excessive reweighting uncertainties.
Apart from increased support as $\sigma_t$ approaches the lower-bound of the hyper-prior, the blue and purple posteriors in Fig.~\ref{fig:flow_gap_posterior} and Fig.~\ref{fig:extra_flow_gap_posterior} appear very similar, indicating that $\approx 5 \times 10^4$ samples per event may be sufficient for this mock problem~\footnote{One can measure the similarity of the approximate posterior to the true posterior (obtained with a large number of samples) using a KL divergence. For the problem studied here, a good rule of thumb seems to be that approximation is qualitatively correct so long as the KL divergence is $\lesssim 0.1$. This condition is met so long as there are $\gtrsim 4\times 10^4$ samples.}.

\section{Discussion}\label{sec:discussion}

Presently, the techniques in Section~\ref{sec:easy_fixes} may reduce the population likelihood's uncertainty enough to safely perform inference on the desired regions of hyper-parameter space.
However, based on the work presented here, we believe that these techniques may become inadequate in the near future when the catalog consists of $\gtrsim 300$ events (potentially after the completion of all O4 data releases).
The issue may arise sooner for population models with sharp features.
In these cases, we can instead quickly bolster PE sample sets via normalizing flows.

Using our mock problem, we show that uncertainty in the population likelihood can be sufficiently reduced even in the case that some events' NPEs are stubbornly inefficient ($\sim1\%$), and rejection sampling efficiency is catastrophically low ($\sim0.1\%$).
In Appendix~\ref{sec:500_event}, we further demonstrate the capabilities of the methods presented here, repeating our demonstrations on an extended mock catalog of 500 events.

Despite our success here, these inefficient NPEs may be a concern for far-future analyses in which we need many more samples from our NPEs, either for use in reweighting, or more advanced population inference techniques (which we discuss further below).
At present, our MCMC-walk solution to low rejection sampling efficiencies is by far the largest bottleneck in our pipeline.
After an NPE has been trained, the process of generating a set of $10^6$ samples weighted to follow the exact posterior takes $\approx 30$ minutes in the slowest cases, but closer to $\approx 5$ minutes for typical events.
Meanwhile, producing unweighted sets of $1 \times 10^5$ samples using our MCMC-walk solution can take up to $\approx 20$ hours.
In other terms, the trained NPEs median importance sampling efficiency of 24\% translates to a cost of $\approx 4$ likelihood evaluations per effective sample.
Meanwhile, our MCMC-walk solution for obtaining unweighted draws increases the median cost to $\approx 100$ evaluations per sample.
While this is still notably better than the nested sampling cost of $\approx 700$ evaluations per posterior sample, we believe there is room for significant improvement.
There are two ways we can foresee this issue being solved.

The first, as alluded to in Section~\ref{sec:flows}, would be to make the importance sampling efficiency of the flows consistently high enough across an entire catalog (heuristically, $\gtrsim50\%$).
This would allow entire batches of flow samples (with weights) to be used directly in population inference via importance sampling, without the need for a prohibitively large number of draws to achieve the desired effective sample size.
As this would eliminate the need to convert the weighted NPE draws into unweighted posterior samples before using them in population inference, the issue of low rejection sampling efficiencies would be bypassed entirely.
As a result, the per-sample cost would be reduced to $\mathcal{O}(1)$ likelihood evaluation for each fiducial sample.
Given we achieved NPE efficiencies of up to 70\%, we see this as plausible with more carefully tailored normalizing flow architectures and extensive training, but we leave exploring this for future work.
The second solution would be eliminating extreme weights in order to deliver an acceptable rejection sampling efficiency (see Section~\ref{sec:flows}).
Achieving a consistent rejection sampling efficiency of anything $>1\%$, would deliver an improvement over our existing per-sample cost of 100 likelihood evaluations for each unweighted posterior draw.
From our investigation of posterior corner plots, the extreme weights often appear somewhere in the bulk of the distribution in one- and two-dimensional slices, potentially exposing pockets of underestimated posterior probabilities in higher dimensions (also see Appendix~\ref{sec:diagnostics}).

To this end, there exists a technique to help mitigate the destructive effects of these extreme weights known as ``defensive sampling'' \citep{Hesterberg:1995, Owen:2000}.
The idea is to artificially inflate the the proposal distribution's (the NPE's) probability across low-density regions of parameter space.
While this hurts the efficiency in the bulk of the distribution, if done correctly, it should reduce the possibility of any sample having a draw probability significantly lower than the target distribution's, thus an extremely large weight (see Eq.~\ref{eq:flow_weight}).
We try several variations of this idea, including sampling from a mixture model between the NPE and the prior (varying from $1\%{-}10\%$ of draws coming from the prior), training the NPE using nested weights modified by an exponent $a < 1$ ($W_i^k \rightarrow {W_i^k}^a$) to smooth and stretch out the distribution, and modifying the Gaussian base distribution of the NPE to have widths $1.05{-}1.2$ times wider than those learned during training.
However, from our brief tests on one of our mock events, this did not result in a notable improvement.
As such, further development of architectures, training techniques or sampling techniques may be necessary in the future.

In Appendix~\ref{sec:diagnostics}, we look for properties shared by events with poor NPE efficiencies.
We find that the worst performing NPEs tend to have lower luminosity distances ($\lesssim 1 \mathrm{Gpc}$).
One might assume this is a result of nearby events having high SNRs, thus sharper structure in the posterior.
However, we find that even low SNR ($\rho \approx 10$) events exhibit poor NPE performance as luminosity distance decreases.
Recalling that we have fit our NPEs to distance marginalized posteriors, it is feasible that the numerical marginalization performed in \textsc{Bilby} (see Refs.~\citep{Ashton:2018jfp, Romero-Shaw:2020owr}) is causing anomalies in the posterior as luminosity distance approaches the lower bound (potentially due to sparseness of the grid over which distance is marginalized), which our normalizing flows struggle to fit.
While amending this could conceivably improve the performance of our worst NPEs, our current implementations prove effective for the demonstrations and aims of this paper, and so we leave exploring this for future work.

As mentioned in Section~\ref{sec:flows}, the \textsc{Dingo} framework \citep{Dax:2022pxd, Dax:2021tsq, Green:2020dnx} is proposed as an alternative to traditional sampling techniques for PE, and may provide an alternative to the NPEs presented here.
\textsc{Dingo} trains a conditional normalizing flow model to generate posterior samples given strain data.
These models are trained for a particular approximant and prior range using $\sim 10^6$ samples from the prior, each coupled with strain data drawn from the likelihood $\mathcal{L}(d|\theta)$.
\textsc{Dingo} solves a more complicated problem than the one presented in Section~\ref{sec:flows}: instead of learning to approximate a single event's posterior using samples directly from the target distribution, a single \textsc{Dingo} model learns to approximate any number of events' posteriors (conditionally on their strain data) using samples from the prior.

Promisingly, \textsc{Dingo} is able to achieve a median efficiency of 10\% on real events from GWTC-3, similar to the efficiencies we achieve on mock events prior to iterative training \cite{Dax:2022pxd}.
On the other hand, \textsc{Dingo} encounters catastrophically low efficiencies for some events ($\sim0.1\%$), attributing these to data quality issues, as well as posteriors being too close to the lower-bound of their model's mass prior \citep{Dax:2022pxd}.
These instances might continue to necessitate PE being performed via traditional sampling algorithms.
In this case, in order to create efficient NPEs for complete catalogs, training on existing posterior samples may prove more reliable than learning unseen posteriors from mock noise and signals.
Our broader point, though, remains that representing single-event posteriors with flows is useful for applications in hierarchical inference---regardless of whether that flow is derived from posterior samples or raw strain data.

\begin{figure}
    \centering
    \includegraphics[width=0.9\columnwidth]{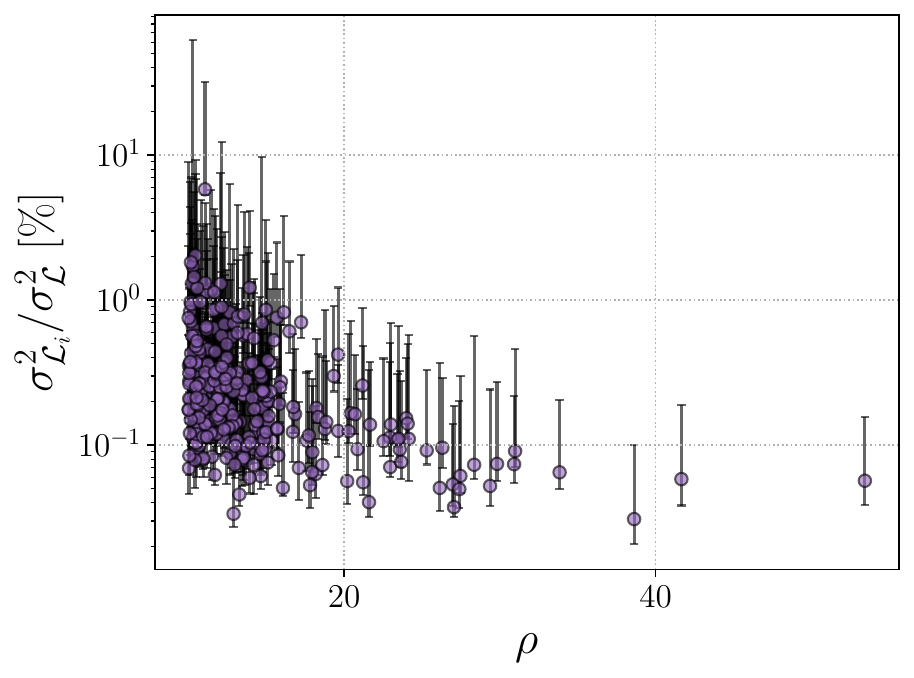}
    \caption{
    Plot of each mock event's relative contribution to the likelihood variance $\sigma_\mathcal{L}^2$ (as a percentage; vertical) against its matched-filter SNR (horizontal).
    The likelihood variances are taken from the population inference in Section~\ref{sec:flows}, using $1\times10^5$ samples to represent each event.
    The circles represent each event's median relative contribution to $\sigma_\mathcal{L}^2$ across all hyper-posterior samples, while the bars represent the maximum and minimum contributions.
    }
    \label{fig:event_vars}
\end{figure}

To save on computational resources, one may wonder if the variance in the population likelihood could be sufficiently reduced by supplementing only some events with additional samples (the ones that contribute most to the likelihood variance in Eq.~\ref{eq:population_uncertainty}), rather than adding samples for every event in the catalog.
To investigate this, we plot each of the 300 events' contributions to the population likelihood variance $\sigma_\mathcal{L}^2$ in Fig.~\ref{fig:event_vars}.
In doing so, we find that some events contribute significantly more to $\sigma_\mathcal{L}^2$ than others, with three of the 300 events in the catalog each occasionally accounting for $> 10\%$ of the likelihood variance, while others consistently account for $< 0.1\%$.
In this mock population, we find the most problematic events do not necessarily sit near sharp features in the mass distribution.
Rather, as Fig.~\ref{fig:event_vars} reveals, events with low SNRs tend to contribute more to $\sigma_\mathcal{L}^2$.
Thus, if adding samples for every event becomes impractical in future catalogs, one might explore two options.
Firstly, as a heuristic approach, the number of samples added to each event could be scaled by its observed SNR.
Alternatively, a more rigorous exploratory analysis could be performed prior to population inference in order to isolate and supplement only the most problematic events.
This exploratory analysis could perhaps consist of sampling from the hyper-prior, then using these hyper-samples to calculate the average contribution to $\sigma_\mathcal{L}^2$ for each event.

Regardless, for analyses of far-future catalogs containing $\gtrsim 10^3$ events, it may become necessary to represent each event with an inordinate number of samples to sufficiently reduce reweighting uncertainty.
At this point, computational constraints will make it impractical to continue relying on contemporary reweighting schemes and new techniques will need to be developed to continue studying the gravitational-wave population.
Normalizing flows are a promising avenue for this.
Ref.~\cite{Mould:2025dts} explore using normalizing flows with variational inference \citep{Blei:2016} to measure gravitational-wave populations.
The authors show that population properties of a catalog consisting of 1599 events can be obtained in minutes by utilizing normalizing flow representations of the hyper-posterior and the associated likelihood variance.
Training the NPE requires orders of magnitude fewer evaluations of the population likelihood than would otherwise be used in traditional population inference \citep{Mould:2025dts}.

Building on the work of Ref.~\cite{Mould:2025dts}, the authors of Ref.~\cite{Wolfe:2026dcq} show that these population-level NPEs can be reused to perform sequential Bayesian updates as gravitational-wave catalogs grow.
The authors show that estimators trained on previous catalogs can be used to avoid repeating population reweighting of past events---meaning a population-level NPE for an updated catalog can be obtained by only reweighting the newly added events.
The methods in Refs.~\citep{Mould:2025dts,Wolfe:2026dcq} are not inherently immune from the variance issues discussed in this work, as the likelihood used to train the population-level NPEs is still calculated via the traditional method of reweighting PE samples.
However, breaking down the reweighting into smaller chunks like Ref.~\cite{Wolfe:2026dcq} may provide the computational headroom to allow the likelihood in Eq.~\ref{eq:loss_mc} to be evaluated with even more samples so that the variance remains negligibly small.

Meanwhile, \textsc{Dingo-Pop} \citep{Leyde:2026hvm} is proposed as an end-to-end population inference pipeline using normalizing flows that, once trained for a single population model, takes $\leq 10^3$ events' (and possibly more in future works) strain data as input and provides an NPE for the resulting population hyper-parameters.
This simulation-based framework was developed from the one introduced in Ref.~\cite{Leyde:2023iof}, which relied on producing PE samples as an intermediate step and required fixing the catalog size prior to training the model.
The authors of Ref.~\cite{Leyde:2026hvm} test \textsc{Dingo-Pop} for a simple power-law and peak mixture model in binary black hole primary mass and mass ratio (see Ref.~\citep{Talbot:2018cva}), finding good agreement with standard hierarchical Bayesian inference methods for a mock catalog of 500 events, and good performance in P-P testing.

Refs.~\citep{Hussain:2025llf, Hussain:2024qzl}, while focusing specifically on under-sampling at the boundaries of parameter space, introduce an entirely different method for overcoming reweighting uncertainties.
Namely, the authors suggest approximating the posteriors (or at least particular dimensions of the posteriors) of gravitational-wave events with mixtures of truncated Gaussians.
If the population model $\pi(\theta|\Lambda)$ is also a mixture of truncated Gaussians (at least in the corresponding dimensions of $\theta$), this allows for Eq.~\ref{eq:population_likelihood} to computed analytically.
This replaces the numerical error from Monte-Carlo integration with the numerical error from approximating PE with truncated Gaussians, which can be substantially smaller and better-behaved under certain circumstances \citep{Hussain:2025llf}.

Finally, the authors of Ref.~\citep{Mancarella:2025uat} present a method to simultaneously infer the source properties $\theta$ of all events in a catalog, along with the population properties $\Lambda$.
This method also eliminates the Monte-Carlo integration seen in Eq.~\ref{eq:population_likelihood_mc}, but again relies on density estimates of individual events' posteriors.
While closely reproducing the inferred population properties of GWTC-3 \citep{KAGRA:2021duu}, it is unclear how this method might scale to significantly larger catalogs.

We believe it will be essential to develop methods for population inference that not only overcome the variance issues presented here while remaining viable for catalogs of $\gtrsim 10^3$ events, but also in which the impact and accuracy of each event's contribution can be carefully traced and vetted.
We suggest that it may be possible to train a neural network to approximate the population likelihood for a single event $\mathcal{L}(d_i|\Lambda)$ with very high accuracy.
This training is embarrassingly parallel, which would enable this pre-processing step being performed efficiently on a high-throughput cluster.

Training such an estimator for each event in the catalog would allow for cheap and precise evaluation of the population likelihood $\mathcal{L}_\mathrm{pop}(d|\Lambda) = \prod_i^N \mathcal{L}(d_i|\Lambda)$ inside of a traditional sampler, with the added benefit of tuning and verifying the accuracy of each event's contribution.
This training would likely require several orders of magnitude more posterior samples for each event than we typically use in gravitational-wave astronomy.
In this case, an efficient NPE for every event's fiducial posterior (potentially using the techniques developed here) would become essential.

As a final point, we note that we have neglected to discuss similar issues pertaining to selection effects $P_\mathrm{det}(\Lambda)$.
Currently, calculating $P_\mathrm{det}(\Lambda)$ also relies on importance sampling (of ``found injections'') meaning that it is subject to the same source of numerical uncertainty explored in this work.
Typically, the variance from this calculation is computed in a similar fashion to Eq.~\ref{eq:population_uncertainty} and added to $\sigma_\mathcal{L}^2$ before applying variance cuts \citep{Talbot:2023pex}.
This means that even if $\sigma_\mathcal{L}^2$ from reweighting PE is well below our threshold, a large variance in $P_\mathrm{det}(\Lambda)$ can reintroduce issues.
Luckily, a number of recent works have explored creating approximations of $P_\mathrm{det}(\Lambda)$ that do not rely on real-time reweighting of fixed injection sets, thus avoiding this limitation \citep{Lorenzo-Medina:2024opt, Callister:2024qyq, Gerosa:2020pgy, Talbot:2020oeu}.
While not done in this work, these techniques could be combined with the methodologies presented here to further eliminate concerns arising from reweighting uncertainties.

\begin{acknowledgments}
We are grateful for our anonymous referee's detailed and insightful review.
We thank Patrick M. Meyers and Liam Pinchbeck for helpful discussions and insights on early iterations of this work.
We also thank Asad Hussain, Matthew Mould and Noah E. Wolfe for insightful comments during the internal LVK review of this manuscript.
This research was supported by the Commonwealth through an Australian Government Research Training Program Scholarship [DOI: \url{https://doi.org/10.82133/C42F-K220}].
The authors acknowledge support from the Australian Research Council (ARC) Centre of Excellence CE170100004, LE210100002, CE230100016,DP230103088 and LE260100008.
This material is based upon work supported by NSF's LIGO Laboratory which is a major facility fully funded by the National Science Foundation.
The authors are grateful for computational resources provided by the LIGO Laboratory and supported by National Science Foundation Grants PHY-0757058 and PHY-0823459.
\end{acknowledgments}

\appendix

\section{Additional demonstrations with a larger mock catalog}\label{sec:500_event}

To show how the methodologies in Sections~\ref{sec:easy_fixes}~and~\ref{sec:flows} scale to larger catalogs, we repeat our analyses on an extended mock catalog of 500 events.
To do so, we first draw an additional 200 mock events and perform PE on them using the exact same methods described in Section~\ref{sec:mock_problem}.
We reuse the same set of $2.5 \times 10^7$ injections for selection effects, finding this still adequately mitigates reweighting uncertainties in $P_\mathrm{det}(\Lambda)$ for a catalog of this size.

\begin{figure*}
    \includegraphics[width=\textwidth]{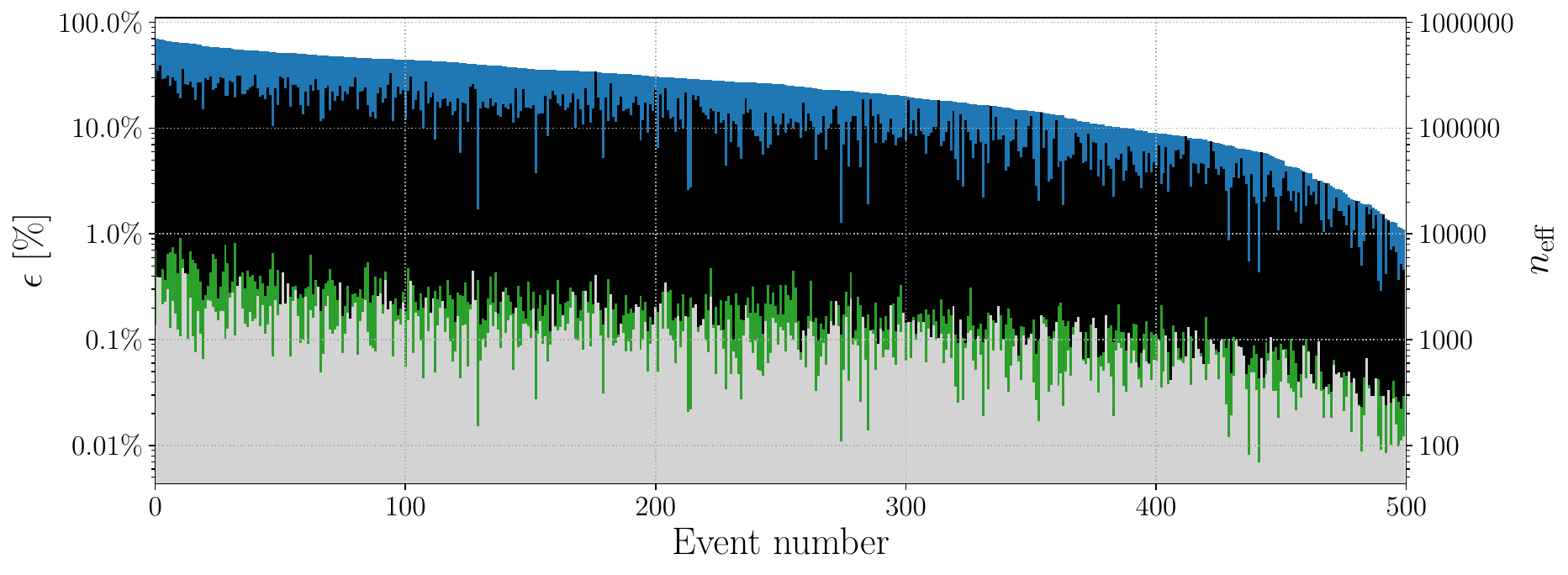}
    \caption{
    Extended bar chart of the normalizing flow performance for each of the 500 mock events.
    As in Fig.~\ref{fig:flow_metrics}, the height of each blue and black bar indicates the effective sample size $n_\mathrm{eff}$ of a batch of $\eta = 10^6$ NPE samples, that have been reweighted to follow the fiducial posterior $p_{\O}(\theta|d_i)$ for the given event.
    This value can also be represented as an efficiency $\epsilon = n_\mathrm{eff}/\eta$, as is displayed on the left axis.
    For each event, the black and blue bars represent the efficiency before and after iterative training respectively.
    Events are numbered and sorted according to the effective sample size achieved by the associated NPE after iterative training (best to worst).
    Meanwhile, the gray and green bars show the number of samples left after rejection sampling draws from the initial NPE and the iteratively trained NPE respectively.
    }
    \label{fig:flow_metrics_500}
\end{figure*}

We construct NPEs for the 200 new events, again following the same steps described in Section~\ref{sec:flows}.
This includes iteratively training all events' NPEs twice, with an additional 2 iterations for the one new stubborn event with an efficiency $\epsilon < 1\%$.
Drawing $\eta=10^6$ samples, we plot a histogram of efficiencies for all 500 events' NPEs (including the initial and iteratively trained) in Fig.~\ref{fig:flow_metrics_500}.
Compared to the initial 300 event catalog, we find similar efficiencies over the extended catalog, with iteratively trained NPEs achieving a median efficiency of 26\% over all 500 events.
As in Section~\ref{sec:flows}, we draw $1 \times 10^5$ unweighted posterior samples from the 200 new NPEs using our redraw and MCMC-walk method.

\begin{figure*}    \includegraphics[width=0.85\textwidth]{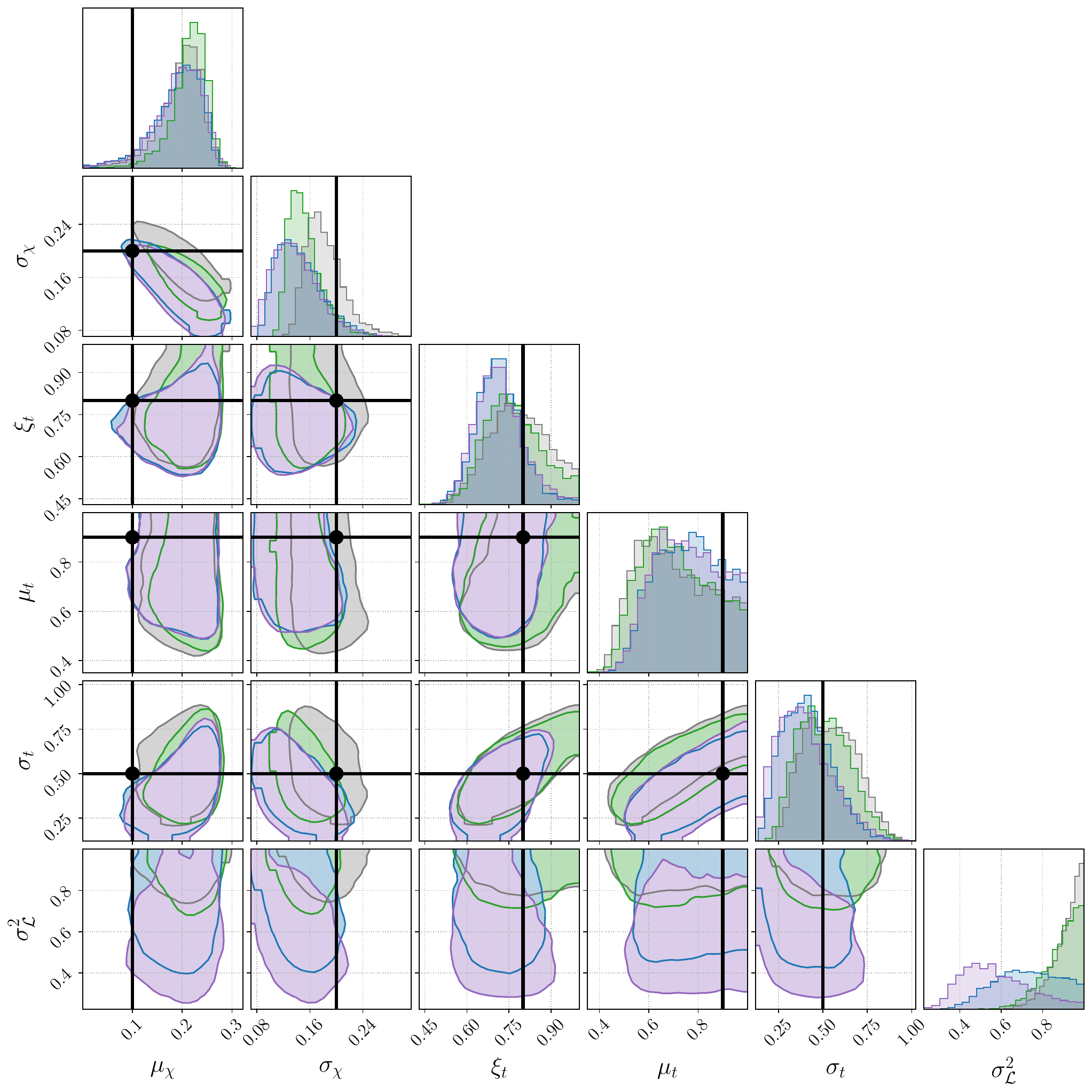}
    \caption{
    Posteriors for spin distribution hyper-parameters, along with the distribution of associated variances in the population likelihood $\sigma_\mathcal{L}^2$, given an extended catalog of 500 mock events.
    Gray shows the hyper-posterior inferred using our original $\approx 1\times 10^4{-}2 \times 10^4$ posterior samples for each event.
    Green uses the entire batch of weighted nested samples produced during PE.
    Meanwhile, blue and purple show the hyper-posteriors inferred using sample sets supplemented with new draws from our NPEs so that each event is represented with $5 \times 10^4$ and $1 \times 10^5$ posterior samples respectively.
    The shaded areas in the two-dimensional panels indicate regions of 90\% credibility.
    The true injected value is shown in black.
    We see little change between the hyper-posteriors inferred with the original posterior samples (gray) and their associated nested samples (green), with both producing variances which rail against the threshold of $\sigma_\mathcal{L}^2=1$.
    Thus, both of these analyses appear insufficient for this catalog.
    Interestingly, the hyper-posteriors inferred with varying numbers of NPE samples (blue and purple) look qualitatively very similar, despite the analysis using only $5 \times 10^4$ draws (blue) producing a relatively large number of hyper-posterior samples with variances close to the threshold.
    The analysis with $1 \times 10^5$ fiducial samples (purple) appears to have converged to the true hyper-posterior given the distribution of $\sigma_\mathcal{L}^2$ values.
    }
    \label{fig:500_spin_posteriors}
\end{figure*}

We then perform population inference on this extended catalog four times.
First, we use the $1\times10^4{-}2\times10^4$ posterior samples obtained via PE.
We then use the $3\times10^4{-}1\times10^5$ weighted nested samples produced during PE (as in Section~\ref{sec:easy_fixes}).
Finally, we perform two analyses in which we supplement our original posterior samples with new draws from our NPEs, one using a total of $5\times10^4$ samples for each event, and one using $1\times10^5$ samples for each event (as in Section~\ref{sec:flows}).
We plot the resulting spin hyper-posteriors and the associated distributions of likelihood variances in Fig.~\ref{fig:500_spin_posteriors}.

We find that using the entire set of nested samples in this case makes a relatively small difference to the hyper-posteriors, with the associated distribution of $\sigma_\mathcal{L}^2$ still peaking sharply at the threshold.
Meanwhile, using $5\times10^4$ samples for each event (produced in part by our NPEs), now produces a large number of variances near the threshold $\sigma_\mathcal{L}^2=1$, and may no longer be sufficient for population inference if we base our assessment off this metric alone.
On the other hand, using $1\times10^5$ samples produces a distribution of variances that peaks well away from the threshold, indicating these results are numerically reliable.
Promisingly, the hyper-posteriors inferred using $5\times10^4$ and $1\times10^5$ posterior samples appear almost identical, indicating that a distribution of variances peaking far from the threshold may not be a requirement for reliable population inference.

\section{Diagnosing NPE performance}\label{sec:diagnostics}

In this section, we perform several diagnostic tests to explore where and why our NPEs underperform.
In these tests, we use the extended catalog of 500 events from Appendix~\ref{sec:500_event}.

\begin{figure}
    \centering
    \includegraphics[width=\columnwidth]{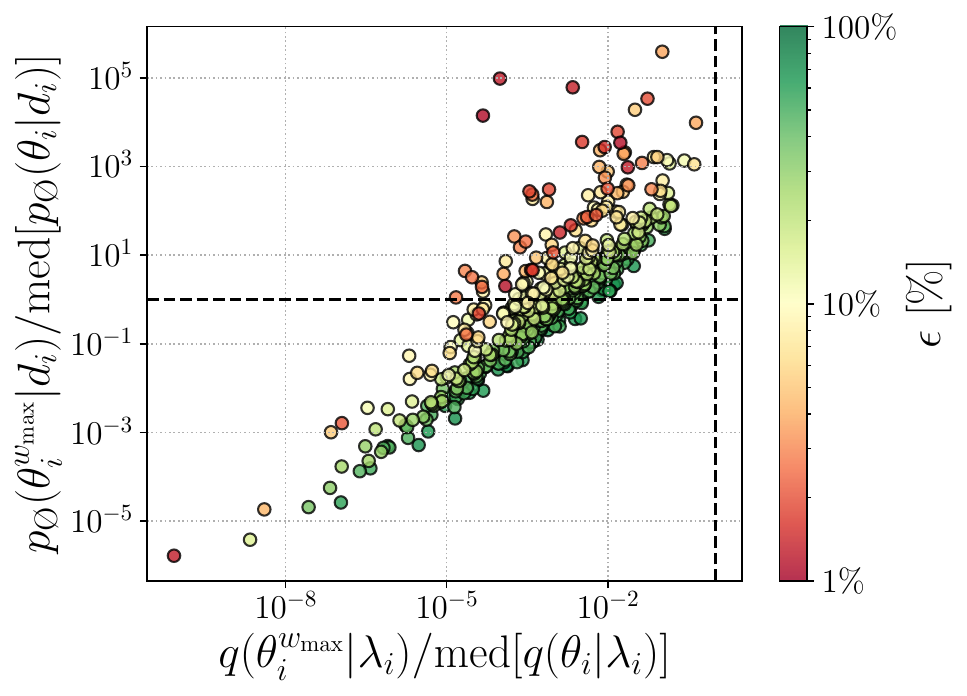}
    \caption{
    A scatter plot of the highest-weight samples $\theta_i^{w_{\max}}$ from each of the 500 events' iteratively trained NPEs.
    The vertical axis displays the true fiducial posterior probability for the highest weight NPE sample relative to the median posterior probability across the entire batch it originates from.
    This can be thought of as a rescaled version of the numerator in the weight (Eq.~\ref{eq:flow_weight}).
    The horizontal axis then represents the denominator in the weight.
    It is the NPE draw probability of the highest weight sample, relative to the median NPE draw probability from the corresponding batch.
    Each sample is color-coded according to the overall efficiency $\epsilon$ (or effective sample size) of the batch of samples it originates from.
    The dashed black lines indicate a ratio of $1$.
    Therefore, samples above (below) these lines indicate posterior or NPE probabilities larger (smaller) than that of a typical sample.
    }
    \label{fig:extreme_weights_numerator_denominator}
\end{figure}

As discussed in Sections~\ref{sec:flows}~and~\ref{sec:discussion}, all of our NPEs produce at least one extreme weight that is $\gtrsim 100$ times larger than that of the corresponding batch's average sample.
To explore the cause of these extreme weights, we isolate the maximum-weight sample $\theta_i^{w_{\max}}$ from each of the 500 event's $\eta = 10^6$ flow samples (drawn from the best-fit iteratively-trained NPE).
We then calculate its fiducial posterior probability relative to that of the median $p_{\O}(\theta_i^{w_{\max}}|d_i) / \mathrm{med}[p_{\O}(\theta_i|d_i)]$, and its NPE probability relative to that of the median $q(\theta_i^{w_{\max}}|\lambda_i) / \mathrm{med}[q(\theta_i|\lambda_i)]$.
These correspond to the numerator's and denominator's contributions to the weights (Eq.~\ref{eq:flow_weight}) relative to that of a typical sample in a given batch.
We plot these in Fig.~\ref{fig:extreme_weights_numerator_denominator}.
The samples are color-coded according to the reweighting efficiency (or effective sample size) of the batch of draws they originate from.
We see a consistent off-diagonal trend in the colors, indicating that the maximum weight relative to the median weight is a reliable indication of the entire batch's effective sample size.

In any batch of samples the most extreme weight consistently has a lower NPE probability (smaller denominator in Eq.~\ref{eq:flow_weight}) than that of the average sample (every point is to the left of the dashed vertical line).
Meanwhile, these extreme-weight samples have posterior probabilities that cluster around that of the average sample.
This trend indicates that the extreme weights in any given batch of samples consistently correspond to under-dense pockets in the NPE, and are not specific to over-dense or under-dense regions of the target posterior.

\begin{figure}
    \centering
    \includegraphics[width=\columnwidth]{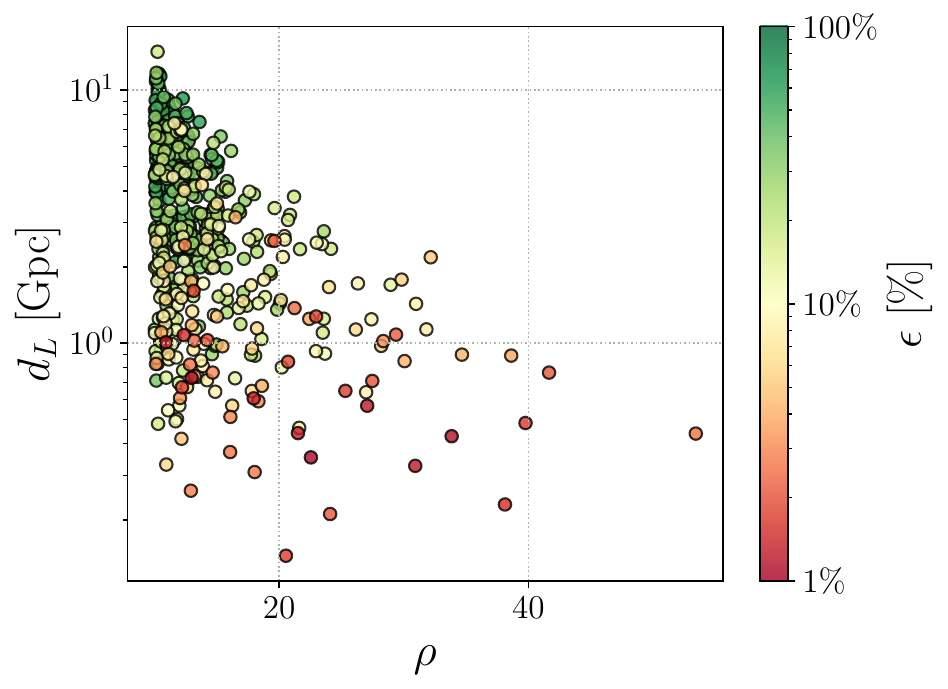}
    \caption{
    A scatter plot for each mock event's injected luminosity distance (vertical) against matched-filter SNR (horizontal), color-coded according to the event's best NPE efficiency.
    The worst NPEs (red) seem to have consistently low luminosity distances, but do not necessarily have high SNRs.
    }
    \label{fig:efficiency_distance_snr}
\end{figure}

In Fig.~\ref{fig:efficiency_distance_snr}, we show a scatter plot for all 500 mock events' injected luminosity distances against their matched-filter SNRs, colored according to the corresponding NPEs' reweighting efficiencies $\epsilon$.
We find that the worst-performing NPEs (lowest efficiencies) tend to have low luminosity distances.
Based on this alone, one might assume this is a byproduct of lower distances producing higher SNRs, thus sharper posteriors.
However, we do not see a clear trend between SNR and NPE efficiency, as even low-SNR ($\rho \approx 10$) events with low luminosity distances tend to have poor NPE performance.
We do not find any clear correlations between any other injected parameters and the performance of the corresponding events' NPEs.
This is with the exception of mass, which is directly tied to luminosity distance through SNR or detectability (low mass events are only detectable at low distances).

Recalling that we have fit our NPEs to distance marginalized posteriors, one possible explanation is that this marginalization is creating anomalies in the posterior when the event's distance approaches the lower bound.
If this is the case, the default-generated grid of distances over which the posterior is marginalized in \textsc{Bilby} may be too sparse for small distances.
While amending this may substantially improve the performance of our worst NPEs, our current implementations prove effective for the demonstrations in this paper, and so we leave exploring this for future work.

\bibliography{refs}

@article{Reitze:2019iox,
    author = "Reitze, David and others",
    title = "{Cosmic Explorer: The U.S. Contribution to Gravitational-Wave Astronomy beyond LIGO}",
    eprint = "1907.04833",
    archivePrefix = "arXiv",
    primaryClass = "astro-ph.IM",
    reportNumber = "LIGO-P1900316",
    journal = "Bull. Am. Astron. Soc.",
    volume = "51",
    number = "7",
    pages = "035",
    year = "2019"
}

@article{Thrane:2018qnx,
    author = "Thrane, Eric and Talbot, Colm",
    title = "{An introduction to Bayesian inference in gravitational-wave astronomy: parameter estimation, model selection, and hierarchical models}",
    eprint = "1809.02293",
    archivePrefix = "arXiv",
    primaryClass = "astro-ph.IM",
    doi = "10.1017/pasa.2019.2",
    journal = "Publ. Astron. Soc. Austral.",
    volume = "36",
    pages = "e010",
    year = "2019",
    note = "[Erratum: Publ.Astron.Soc.Austral. 37, e036 (2020)]"
}

@article{Talbot:2023pex,
    author = "Talbot, Colm and Golomb, Jacob",
    title = "{Growing pains: understanding the impact of likelihood uncertainty on hierarchical Bayesian inference for gravitational-wave astronomy}",
    eprint = "2304.06138",
    archivePrefix = "arXiv",
    primaryClass = "astro-ph.IM",
    doi = "10.1093/mnras/stad2968",
    journal = "Mon. Not. Roy. Astron. Soc.",
    volume = "526",
    number = "3",
    pages = "3495--3503",
    year = "2023"
}

@article{Callister:2024cdx,
    author = "Callister, T. A.",
    title = "{Observed Gravitational-Wave Populations}",
    eprint = "2410.19145",
    archivePrefix = "arXiv",
    primaryClass = "astro-ph.HE",
    month = "10",
    year = "2024"
}

@article{Lorenzo-Medina:2024opt,
    author = "Lorenzo-Medina, Ana and Dent, Thomas",
    title = "{A physically modelled selection function for compact binary mergers in the LIGO-Virgo O3 run and beyond}",
    eprint = "2408.13383",
    archivePrefix = "arXiv",
    primaryClass = "gr-qc",
    doi = "10.1088/1361-6382/ad9c0e",
    journal = "Class. Quant. Grav.",
    volume = "42",
    number = "4",
    pages = "045008",
    year = "2025"
}

@article{Callister:2024qyq,
    author = "Callister, Thomas A. and Essick, Reed and Holz, Daniel E.",
    title = "{Neural network emulator of the Advanced LIGO and Advanced Virgo selection function}",
    eprint = "2408.16828",
    archivePrefix = "arXiv",
    primaryClass = "astro-ph.HE",
    doi = "10.1103/PhysRevD.110.123041",
    journal = "Phys. Rev. D",
    volume = "110",
    number = "12",
    pages = "123041",
    year = "2024"
}

@article{Gerosa:2020pgy,
    author = "Gerosa, Davide and Pratten, Geraint and Vecchio, Alberto",
    title = "{Gravitational-wave selection effects using neural-network classifiers}",
    eprint = "2007.06585",
    archivePrefix = "arXiv",
    primaryClass = "astro-ph.HE",
    doi = "10.1103/PhysRevD.102.103020",
    journal = "Phys. Rev. D",
    volume = "102",
    number = "10",
    pages = "103020",
    year = "2020"
}

@article{Gair:2022zsa,
    author = "Gair, Jonathan R. and others",
    title = "{The Hitchhiker{\textquoteright}s Guide to the Galaxy Catalog Approach for Dark Siren Gravitational-wave Cosmology}",
    eprint = "2212.08694",
    archivePrefix = "arXiv",
    primaryClass = "gr-qc",
    doi = "10.3847/1538-3881/acca78",
    journal = "Astron. J.",
    volume = "166",
    number = "1",
    pages = "22",
    year = "2023"
}

@article{Frazier:2024,
       author = {{Frazier}, David T. and {Kelly}, Ryan and {Drovandi}, Christopher and {Warne}, David J.},
        title = "{The Statistical Accuracy of Neural Posterior and Likelihood Estimation}",
         year = 2024,
        month = 11,
archivePrefix = {arXiv},
       eprint = {2411.12068},
 primaryClass = {stat.ML},
}

@article{Gilks:2001,
    author = {Gilks, Walter R. and Berzuini, Carlo},
    title = {Following a Moving Target—Monte Carlo Inference for Dynamic Bayesian Models},
    journal = {Journal of the Royal Statistical Society Series B: Statistical Methodology},
    volume = {63},
    number = {1},
    pages = {127-146},
    year = {2001},
    month = {1},
    issn = {1369-7412},
    doi = {10.1111/1467-9868.00280},
}

@article{Doucet:2009,
    author = {Doucet, Arnaud and Johansen, Adam},
    year = {2009},
    month = {01},
    pages = {656-704},
    title = {A Tutorial on Particle Filtering and Smoothing: Fifteen Years Later},
    volume = {12},
    ISBN = {9780199532902},
    journal = {Handbook of Nonlinear Filtering}
}

@article{Galaudage:2024meo,
    author = "Galaudage, Shanika and Lamberts, Astrid",
    title = "{Compactness peaks: An astrophysical interpretation of the mass distribution of merging binary black holes}",
    eprint = "2407.17561",
    archivePrefix = "arXiv",
    primaryClass = "astro-ph.HE",
    doi = "10.1051/0004-6361/202451654",
    journal = "Astron. Astrophys.",
    volume = "694",
    pages = "A186",
    year = "2025"
}

@article{Legred:2026oiz,
    author = "Legred, Isaac and Golomb, Jacob and Chatziioannou, Katerina",
    title = "{Low-mass failed supernovae and the $10\,M_{\odot}$ peak in the merging black hole mass distribution}",
    eprint = "2604.01420",
    archivePrefix = "arXiv",
    primaryClass = "astro-ph.HE",
    month = "4",
    year = "2026"
}

@article{LIGOScientific:2026qni,
    author = "Abac, A. G. and others",
    collaboration = "LIGO Scientific, VIRGO, KAGRA",
    title = "{GWTC-4.0: Tests of General Relativity. I. Overview and General Tests}",
    eprint = "2603.19019",
    archivePrefix = "arXiv",
    primaryClass = "gr-qc",
    reportNumber = "LIGO-P2500065",
    month = "3",
    year = "2026"
}

@article{LIGOScientific:2026fcf,
    author = "Abac, A. G. and others",
    collaboration = "LIGO Scientific, VIRGO, KAGRA",
    title = "{GWTC-4.0: Tests of General Relativity. II. Parameterized Tests}",
    eprint = "2603.19020",
    archivePrefix = "arXiv",
    primaryClass = "gr-qc",
    reportNumber = "LIGO-P2500066",
    month = "3",
    year = "2026"
}

@article{Farr:2019rap,
    author = "Farr, Will M.",
    title = "{Accuracy Requirements for Empirically-Measured Selection Functions}",
    eprint = "1904.10879",
    archivePrefix = "arXiv",
    primaryClass = "astro-ph.IM",
    doi = "10.3847/2515-5172/ab1d5f",
    journal = "Research Notes of the AAS",
    volume = "3",
    number = "5",
    pages = "66",
    year = "2019"
}

@article{Essick:2022ojx,
    author = "Essick, Reed and Farr, Will",
    title = "{Precision Requirements for Monte Carlo Sums within Hierarchical Bayesian Inference}",
    eprint = "2204.00461",
    archivePrefix = "arXiv",
    primaryClass = "astro-ph.IM",
    month = "4",
    year = "2022"
}

@article{LIGOScientific:2025pvj,
    author = "Abac, A. G. and others",
    collaboration = "LIGO Scientific, VIRGO, KAGRA",
    title = "{GWTC-4.0: Population Properties of Merging Compact Binaries}",
    eprint = "2508.18083",
    archivePrefix = "arXiv",
    primaryClass = "astro-ph.HE",
    reportNumber = "LIGO-P2400004",
    month = "8",
    year = "2025"
}

@article{LIGOScientific:2025jau,
    author = "Abac, A. G. and others",
    collaboration = "LIGO Scientific, VIRGO, KAGRA",
    title = "{GWTC-4.0: Constraints on the Cosmic Expansion Rate and Modified Gravitational-wave Propagation}",
    eprint = "2509.04348",
    archivePrefix = "arXiv",
    primaryClass = "astro-ph.CO",
    reportNumber = "LIGO-P2400152",
    month = "9",
    year = "2025"
}

@article{Mandel:2018mve,
    author = "Mandel, Ilya and Farr, Will M. and Gair, Jonathan R.",
    title = "{Extracting distribution parameters from multiple uncertain observations with selection biases}",
    eprint = "1809.02063",
    archivePrefix = "arXiv",
    primaryClass = "physics.data-an",
    doi = "10.1093/mnras/stz896",
    journal = "Mon. Not. Roy. Astron. Soc.",
    volume = "486",
    number = "1",
    pages = "1086--1093",
    year = "2019"
}

@article{Tiwari:2017ndi,
    author = "Tiwari, Vaibhav",
    title = "{Estimation of the Sensitive Volume for Gravitational-wave Source Populations Using Weighted Monte Carlo Integration}",
    eprint = "1712.00482",
    archivePrefix = "arXiv",
    primaryClass = "astro-ph.HE",
    doi = "10.1088/1361-6382/aac89d",
    journal = "Class. Quant. Grav.",
    volume = "35",
    number = "14",
    pages = "145009",
    year = "2018"
}

@article{Adamcewicz:2024jkr,
    author = "Adamcewicz, Christian and Lasky, Paul D. and Thrane, Eric and Mandel, Ilya",
    title = "{No Evidence for a Dip in the Binary Black Hole Mass Spectrum}",
    eprint = "2406.11111",
    archivePrefix = "arXiv",
    primaryClass = "astro-ph.HE",
    doi = "10.3847/1538-4357/ad7ea8",
    journal = "Astrophys. J.",
    volume = "975",
    number = "2",
    pages = "253",
    year = "2024"
}

@article{Fishbach:2018edt,
    author = "Fishbach, Maya and Holz, Daniel E. and Farr, Will M.",
    title = "{Does the Black Hole Merger Rate Evolve with Redshift?}",
    eprint = "1805.10270",
    archivePrefix = "arXiv",
    primaryClass = "astro-ph.HE",
    doi = "10.3847/2041-8213/aad800",
    journal = "Astrophys. J. Lett.",
    volume = "863",
    number = "2",
    pages = "L41",
    year = "2018"
}

@article{Vitale:2022dpa,
    author = "Vitale, Salvatore and Biscoveanu, Sylvia and Talbot, Colm",
    title = "{Spin it as you like: The (lack of a) measurement of the spin tilt distribution with LIGO-Virgo-KAGRA binary black holes}",
    eprint = "2209.06978",
    archivePrefix = "arXiv",
    primaryClass = "astro-ph.HE",
    doi = "10.1051/0004-6361/202245084",
    journal = "Astron. Astrophys.",
    volume = "668",
    pages = "L2",
    year = "2022"
}

@article{Talbot:2017yur,
    author = "Talbot, Colm and Thrane, Eric",
    title = "{Determining the population properties of spinning black holes}",
    eprint = "1704.08370",
    archivePrefix = "arXiv",
    primaryClass = "astro-ph.HE",
    doi = "10.1103/PhysRevD.96.023012",
    journal = "Phys. Rev. D",
    volume = "96",
    number = "2",
    pages = "023012",
    year = "2017"
}

@article{Schneider:2023mxe,
    author = "Schneider, Fabian R. N. and Podsiadlowski, Philipp and Laplace, Eva",
    title = "{Bimodal Black Hole Mass Distribution and Chirp Masses of Binary Black Hole Mergers}",
    eprint = "2305.02380",
    archivePrefix = "arXiv",
    primaryClass = "astro-ph.HE",
    doi = "10.3847/2041-8213/acd77a",
    journal = "Astrophys. J. Lett.",
    volume = "950",
    number = "2",
    pages = "L9",
    year = "2023"
}

@article{Talbot:2018cva,
    author = "Talbot, Colm and Thrane, Eric",
    title = "{Measuring the binary black hole mass spectrum with an astrophysically motivated parameterization}",
    eprint = "1801.02699",
    archivePrefix = "arXiv",
    primaryClass = "astro-ph.HE",
    doi = "10.3847/1538-4357/aab34c",
    journal = "Astrophys. J.",
    volume = "856",
    number = "2",
    pages = "173",
    year = "2018"
}

@article{Schneider:2020vvh,
    author = {Schneider, F. R. N. and Podsiadlowski, Ph. and M{\"u}ller, B.},
    title = "{Pre-supernova evolution, compact object masses and explosion properties of stripped binary stars}",
    eprint = "2008.08599",
    archivePrefix = "arXiv",
    primaryClass = "astro-ph.SR",
    doi = "10.1051/0004-6361/202039219",
    journal = "Astron. Astrophys.",
    volume = "645",
    pages = "A5",
    year = "2021"
}

@article{Adamcewicz:2025phm,
    author = "Adamcewicz, Christian and Guttman, Nir and Lasky, Paul D. and Thrane, Eric",
    title = "{Do Both Black Holes Spin in Merging Binaries? Evidence from GWTC-4 and Astrophysical Implications}",
    eprint = "2509.04706",
    archivePrefix = "arXiv",
    primaryClass = "astro-ph.HE",
    doi = "10.3847/1538-4357/ae1370",
    journal = "Astrophys. J.",
    volume = "994",
    number = "2",
    pages = "261",
    year = "2025"
}

@article{Adamcewicz:2023szp,
    author = "Adamcewicz, Christian and Galaudage, Shanika and Lasky, Paul D. and Thrane, Eric",
    title = "{Which Black Hole Is Spinning? Probing the Origin of Black Hole Spin with Gravitational Waves}",
    eprint = "2311.05182",
    archivePrefix = "arXiv",
    primaryClass = "astro-ph.HE",
    doi = "10.3847/2041-8213/ad2df2",
    journal = "Astrophys. J. Lett.",
    volume = "964",
    number = "1",
    pages = "L6",
    year = "2024"
}

@article{Tong:2022iws,
    author = "Tong, Hui and Galaudage, Shanika and Thrane, Eric",
    title = "{Population properties of spinning black holes using the gravitational-wave transient catalog 3}",
    eprint = "2209.02206",
    archivePrefix = "arXiv",
    primaryClass = "astro-ph.HE",
    doi = "10.1103/PhysRevD.106.103019",
    journal = "Phys. Rev. D",
    volume = "106",
    number = "10",
    pages = "103019",
    year = "2022"
}

@article{Galaudage:2021rkt,
    author = "Galaudage, Shanika and others",
    title = "{Building Better Spin Models for Merging Binary Black Holes: Evidence for Nonspinning and Rapidly Spinning Nearly Aligned Subpopulations}",
    eprint = "2109.02424",
    archivePrefix = "arXiv",
    primaryClass = "gr-qc",
    doi = "10.3847/2041-8213/ac2f3c",
    journal = "Astrophys. J. Lett.",
    volume = "921",
    number = "1",
    pages = "L15",
    year = "2021",
    note = "[Erratum: Astrophys.J.Lett. 936, L18 (2022), Erratum: Astrophys.J. 936, L18 (2022)]"
}

@article{Pratten:2020ceb,
    author = "Pratten, Geraint and others",
    title = "{Computationally efficient models for the dominant and subdominant harmonic modes of precessing binary black holes}",
    eprint = "2004.06503",
    archivePrefix = "arXiv",
    primaryClass = "gr-qc",
    doi = "10.1103/PhysRevD.103.104056",
    journal = "Phys. Rev. D",
    volume = "103",
    number = "10",
    pages = "104056",
    year = "2021"
}

@article{Roulet:2024hwz,
    author = "Roulet, Javier and Mushkin, Jonathan and Wadekar, Digvijay and Venumadhav, Tejaswi and Zackay, Barak and Zaldarriaga, Matias",
    title = "{Fast marginalization algorithm for optimizing gravitational wave detection, parameter estimation, and sky localization}",
    eprint = "2404.02435",
    archivePrefix = "arXiv",
    primaryClass = "gr-qc",
    doi = "10.1103/PhysRevD.110.044010",
    journal = "Phys. Rev. D",
    volume = "110",
    number = "4",
    pages = "044010",
    year = "2024"
}

@article{Ashton:2018jfp,
    author = "Ashton, Gregory and others",
    title = "{BILBY: A user-friendly Bayesian inference library for gravitational-wave astronomy}",
    eprint = "1811.02042",
    archivePrefix = "arXiv",
    primaryClass = "astro-ph.IM",
    doi = "10.3847/1538-4365/ab06fc",
    journal = "Astrophys. J. Suppl.",
    volume = "241",
    number = "2",
    pages = "27",
    year = "2019"
}

@article{Romero-Shaw:2020owr,
    author = "Romero-Shaw, I. M. and others",
    title = "{Bayesian inference for compact binary coalescences with bilby: validation and application to the first LIGO\textendash{}Virgo gravitational-wave transient catalogue}",
    eprint = "2006.00714",
    archivePrefix = "arXiv",
    primaryClass = "astro-ph.IM",
    doi = "10.1093/mnras/staa2850",
    journal = "Mon. Not. Roy. Astron. Soc.",
    volume = "499",
    number = "3",
    pages = "3295--3319",
    year = "2020"
}

@article{Veitch:2014wba,
    author = "Veitch, J. and others",
    title = "{Parameter estimation for compact binaries with ground-based gravitational-wave observations using the LALInference software library}",
    eprint = "1409.7215",
    archivePrefix = "arXiv",
    primaryClass = "gr-qc",
    reportNumber = "LIGO-P1400152",
    doi = "10.1103/PhysRevD.91.042003",
    journal = "Phys. Rev. D",
    volume = "91",
    number = "4",
    pages = "042003",
    year = "2015"
}

@article{Speagle:2019ivv,
    author = "Speagle, Joshua S.",
    title = "{dynesty: a dynamic nested sampling package for estimating Bayesian posteriors and evidences}",
    eprint = "1904.02180",
    archivePrefix = "arXiv",
    primaryClass = "astro-ph.IM",
    doi = "10.1093/mnras/staa278",
    journal = "Mon. Not. Roy. Astron. Soc.",
    volume = "493",
    number = "3",
    pages = "3132--3158",
    year = "2020"
}

@article{LIGOScientific:2025slb,
    author = "Abac, A. G. and others",
    collaboration = "LIGO Scientific, VIRGO, KAGRA",
    title = "{GWTC-4.0: Updating the Gravitational-Wave Transient Catalog with Observations from the First Part of the Fourth LIGO-Virgo-KAGRA Observing Run}",
    eprint = "2508.18082",
    archivePrefix = "arXiv",
    primaryClass = "gr-qc",
    reportNumber = "LIGO-P2400386",
    month = "8",
    year = "2025"
}

@article{Talbot:2019okv,
    author = "Talbot, Colm and Smith, Rory and Thrane, Eric and Poole, Gregory B.",
    title = "{Parallelized Inference for Gravitational-Wave Astronomy}",
    eprint = "1904.02863",
    archivePrefix = "arXiv",
    primaryClass = "astro-ph.IM",
    doi = "10.1103/PhysRevD.100.043030",
    journal = "Phys. Rev. D",
    volume = "100",
    number = "4",
    pages = "043030",
    year = "2019"
}

@article{KAGRA:2013rdx,
    author = "Abbott, B. P. and others",
    collaboration = "KAGRA, LIGO Scientific, Virgo",
    title = "{Prospects for observing and localizing gravitational-wave transients with Advanced LIGO, Advanced Virgo and KAGRA}",
    eprint = "1304.0670",
    archivePrefix = "arXiv",
    primaryClass = "gr-qc",
    reportNumber = "LIGO-P1200087, VIR-0288A-12, JGW-P1808427",
    doi = "10.1007/s41114-020-00026-9",
    journal = "Living Rev. Rel.",
    volume = "19",
    pages = "1",
    year = "2016"
}

@article{Talbot:2024yqw,
    author = "Talbot, Colm and Farah, Amanda and Galaudage, Shanika and Golomb, Jacob and Tong, Hui",
    title = "{GWPopulation: Hardware agnostic population inference for compact binaries and beyond}",
    eprint = "2409.14143",
    archivePrefix = "arXiv",
    primaryClass = "astro-ph.IM",
    doi = "10.21105/joss.07753",
    journal = "J. Open Source Softw.",
    volume = "10",
    number = "109",
    pages = "7753",
    year = "2025"
}

@software{ward2023flowjax,
  title = {FlowJAX: Distributions and Normalizing Flows in Jax},
  author = {Daniel Ward},
  url = {https://github.com/danielward27/flowjax},
  version = {17.2.1},
  year = {2022},
  doi = {10.5281/zenodo.10402073},
}

@inproceedings{Durkan:2019nsq,
    author = "Durkan, Conor and Bekasov, Artur and Murray, Iain and Papamakarios, George",
    title = "{Neural Spline Flows}",
    eprint = "1906.04032",
    archivePrefix = "arXiv",
    primaryClass = "stat.ML",
    month = "6",
    year = "2019"
}

@article{Rezende:2020hrd,
    author = "Rezende, Danilo Jimenez and Papamakarios, George and Racani{\`e}re, S{\'e}bastien and Albergo, Michael S. and Kanwar, Gurtej and Shanahan, Phiala E. and Cranmer, Kyle",
    title = "{Normalizing Flows on Tori and Spheres}",
    eprint = "2002.02428",
    archivePrefix = "arXiv",
    primaryClass = "stat.ML",
    month = "2",
    year = "2020"
}

@article{Heinzel:2025ogf,
    author = "Heinzel, Jack and Vitale, Salvatore",
    title = "{When (not) to trust Monte Carlo approximations for hierarchical Bayesian inference}",
    eprint = "2509.07221",
    archivePrefix = "arXiv",
    primaryClass = "astro-ph.HE",
    month = "9",
    year = "2025"
}

@article{LIGOScientific:2014pky,
    author = "Aasi, J. and others",
    collaboration = "LIGO Scientific",
    title = "{Advanced LIGO}",
    eprint = "1411.4547",
    archivePrefix = "arXiv",
    primaryClass = "gr-qc",
    doi = "10.1088/0264-9381/32/7/074001",
    journal = "Class. Quant. Grav.",
    volume = "32",
    pages = "074001",
    year = "2015"
}

@article{VIRGO:2014yos,
    author = "Acernese, F. and others",
    collaboration = "VIRGO",
    title = "{Advanced Virgo: a second-generation interferometric gravitational wave detector}",
    eprint = "1408.3978",
    archivePrefix = "arXiv",
    primaryClass = "gr-qc",
    doi = "10.1088/0264-9381/32/2/024001",
    journal = "Class. Quant. Grav.",
    volume = "32",
    number = "2",
    pages = "024001",
    year = "2015"
}

@article{KAGRA:2020tym,
    author = "Akutsu, T. and others",
    collaboration = "KAGRA",
    title = "{Overview of KAGRA: Detector design and construction history}",
    eprint = "2005.05574",
    archivePrefix = "arXiv",
    primaryClass = "physics.ins-det",
    doi = "10.1093/ptep/ptaa125",
    journal = "PTEP",
    volume = "2021",
    number = "5",
    pages = "05A101",
    year = "2021"
}

@article{Capote:2024rmo,
    author = "Capote, E. and others",
    title = "{Advanced LIGO detector performance in the fourth observing run}",
    eprint = "2411.14607",
    archivePrefix = "arXiv",
    primaryClass = "gr-qc",
    reportNumber = "LIGO-P2400256",
    doi = "10.1103/PhysRevD.111.062002",
    journal = "Phys. Rev. D",
    volume = "111",
    number = "6",
    pages = "062002",
    year = "2025"
}

@article{LIGO:2024kkz,
    author = "Soni, S. and others",
    collaboration = "LIGO",
    title = "{LIGO Detector Characterization in the first half of the fourth Observing run}",
    eprint = "2409.02831",
    archivePrefix = "arXiv",
    primaryClass = "astro-ph.IM",
    doi = "10.1088/1361-6382/adc4b6",
    journal = "Class. Quant. Grav.",
    volume = "42",
    number = "8",
    pages = "085016",
    year = "2025"
}

@article{aLIGO:2020wna,
    author = "Buikema, Aaron and others",
    collaboration = "aLIGO",
    title = "{Sensitivity and performance of the Advanced LIGO detectors in the third observing run}",
    eprint = "2008.01301",
    archivePrefix = "arXiv",
    primaryClass = "astro-ph.IM",
    doi = "10.1103/PhysRevD.102.062003",
    journal = "Phys. Rev. D",
    volume = "102",
    number = "6",
    pages = "062003",
    year = "2020"
}

@article{Prathaban:2026kft,
    author = "Prathaban, Metha and Hoy, Charlie and Williams, Michael J.",
    title = "{Leveraging rapid parameter estimates for efficient gravitational-wave Bayesian inference via posterior repartitioning}",
    eprint = "2601.21630",
    archivePrefix = "arXiv",
    primaryClass = "gr-qc",
    reportNumber = "LIGO-P2600020",
    month = "1",
    year = "2026"
}

@article{Fairhurst:2023idl,
    author = "Fairhurst, Stephen and Hoy, Charlie and Green, Rhys and Mills, Cameron and Usman, Samantha A.",
    title = "{Simple parameter estimation using observable features of gravitational-wave signals}",
    eprint = "2304.03731",
    archivePrefix = "arXiv",
    primaryClass = "gr-qc",
    doi = "10.1103/PhysRevD.108.082006",
    journal = "Phys. Rev. D",
    volume = "108",
    number = "8",
    pages = "082006",
    year = "2023"
}

@article{Mould:2025dts,
    author = "Mould, Matthew and Wolfe, Noah E. and Vitale, Salvatore",
    title = "{Rapid inference and comparison of gravitational-wave population models with neural variational posteriors}",
    eprint = "2504.07197",
    archivePrefix = "arXiv",
    primaryClass = "astro-ph.IM",
    doi = "10.1103/xk1z-fxnm",
    journal = "Phys. Rev. D",
    volume = "111",
    number = "12",
    pages = "123049",
    year = "2025"
}

@article{Wolfe:2026dcq,
    author = "Wolfe, Noah E. and Mould, Matthew and Veitch, John and Vitale, Salvatore",
    title = "{Neural Bayesian updates to populations with growing gravitational-wave catalogs}",
    eprint = "2602.20277",
    archivePrefix = "arXiv",
    primaryClass = "astro-ph.IM",
    month = "2",
    year = "2026"
}

@article{Leyde:2023iof,
    author = "Leyde, Konstantin and Green, Stephen R. and Toubiana, Alexandre and Gair, Jonathan",
    title = "{Gravitational wave populations and cosmology with neural posterior estimation}",
    eprint = "2311.12093",
    archivePrefix = "arXiv",
    primaryClass = "gr-qc",
    doi = "10.1103/PhysRevD.109.064056",
    journal = "Phys. Rev. D",
    volume = "109",
    number = "6",
    pages = "064056",
    year = "2024"
}

@article{Hussain:2025llf,
    author = "Hussain, Asad and Isi, Maximiliano and Zimmerman, Aaron",
    title = "{Living on the edge: Testing for compact population features at the edges of parameter space}",
    eprint = "2510.20010",
    archivePrefix = "arXiv",
    primaryClass = "astro-ph.IM",
    month = "10",
    year = "2025"
}

@article{Hussain:2024qzl,
    author = "Hussain, Asad and Isi, Maximiliano and Zimmerman, Aaron",
    title = "{Hints of Spin-magnitude Correlations and a Rapidly Spinning Subpopulation of Binary Black Holes}",
    eprint = "2411.02252",
    archivePrefix = "arXiv",
    primaryClass = "astro-ph.HE",
    reportNumber = "LIGO-P2400522, UTWI-33-2024",
    doi = "10.3847/1538-4357/ae1574",
    journal = "Astrophys. J.",
    volume = "996",
    number = "1",
    pages = "71",
    year = "2026"
}

@article{Williams:2025aar,
    author = "Williams, Michael J.",
    title = "{Accelerated Sequential Posterior Inference via Reuse for Gravitational-Wave Analyses}",
    eprint = "2511.04218",
    archivePrefix = "arXiv",
    primaryClass = "hep-ex",
    reportNumber = "LIGO-P2500623",
    month = "11",
    year = "2025"
}

@article{Leyde:2026hvm,
    author = "Leyde, Konstantin and Green, Stephen R. and Dax, Maximilian and Mould, Matthew and Fabbri, Cecilia Maria and Gair, Jonathan",
    title = "{End-to-End Population Inference from Gravitational-Wave Strain using Transformers}",
    eprint = "2605.11274",
    archivePrefix = "arXiv",
    primaryClass = "gr-qc",
    month = "5",
    year = "2026"
}

@article{Prathaban:2024rmu,
    author = "Prathaban, Metha and Bevins, Harry and Handley, Will",
    title = "{Accelerated nested sampling with posterior repartitioning and {\ensuremath{\beta}}-flows for gravitational waves}",
    eprint = "2411.17663",
    archivePrefix = "arXiv",
    primaryClass = "astro-ph.IM",
    doi = "10.1093/mnras/staf962",
    journal = "Mon. Not. Roy. Astron. Soc.",
    volume = "541",
    number = "1",
    pages = "200--213",
    year = "2025"
}

@article{Hesterberg:1995,
    author = {Tim Hesterberg},
    title = {Weighted Average Importance Sampling and Defensive Mixture Distributions},
    journal = {Technometrics},
    volume = {37},
    number = {2},
    pages = {185--194},
    year = {1995},
    doi = {10.1080/00401706.1995.10484303},
}

@article{Owen:2000,
    author = {Art Owen and Yi Zhou Associate},
    title = {Safe and Effective Importance Sampling},
    journal = {Journal of the American Statistical Association},
    volume = {95},
    number = {449},
    pages = {135--143},
    year = {2000},
    doi = {10.1080/01621459.2000.10473909},
}

@article{Papamakarios:2019fms,
    author = "Papamakarios, George and Nalisnick, Eric and Rezende, Danilo Jimenez and Mohamed, Shakir and Lakshminarayanan, Balaji",
    title = "{Normalizing Flows for Probabilistic Modeling and Inference}",
    eprint = "1912.02762",
    archivePrefix = "arXiv",
    primaryClass = "stat.ML",
    doi = "10.5555/3546258.3546315",
    journal = "J. Machine Learning Res.",
    volume = "22",
    number = "1",
    pages = "2617--2680",
    year = "2021"
}

@article{Skilling:2006gxv,
    author = "Skilling, John",
    title = "{Nested sampling for general Bayesian computation}",
    doi = "10.1214/06-BA127",
    journal = "Bayesian Analysis",
    volume = "1",
    number = "4",
    pages = "833--859",
    year = "2006"
}

@article{LIGOO4Detector:2023wmz,
    author = "Ganapathy, D. and others",
    collaboration = "LIGO O4 Detector",
    title = "{Broadband Quantum Enhancement of the LIGO Detectors with Frequency-Dependent Squeezing}",
    doi = "10.1103/PhysRevX.13.041021",
    journal = "Phys. Rev. X",
    volume = "13",
    number = "4",
    pages = "041021",
    year = "2023"
}

@article{membersoftheLIGOScientific:2024elc,
    author = "Jia, Wenxuan and others",
    collaboration = "members of the LIGO Scientific{\textdagger}",
    title = "{Squeezing the quantum noise of a gravitational-wave detector below the standard quantum limit}",
    eprint = "2404.14569",
    archivePrefix = "arXiv",
    primaryClass = "gr-qc",
    reportNumber = "LIGO-P2400059",
    doi = "10.1126/science.ado8069",
    journal = "Science",
    volume = "385",
    number = "6715",
    pages = "1318",
    year = "2024"
}

@article{Blei:2016,
    author = {{Blei}, David M. and {Kucukelbir}, Alp and {McAuliffe}, Jon D.},
    title = "{Variational Inference: A Review for Statisticians}",
    eprint = "1601.00670",
    archivePrefix = "arXiv",
    primaryClass = {stat.CO},
    doi = "10.1080/01621459.2017.1285773",
    journal = {Journal of the American Statistical Association},
    volume = "112",
    number = "518",
    pages = "859--877",
    year = "2017"
}

@article{KAGRA:2021duu,
    author = "Abbott, R. and others",
    collaboration = "KAGRA, VIRGO, LIGO Scientific",
    title = "{Population of Merging Compact Binaries Inferred Using Gravitational Waves through GWTC-3}",
    eprint = "2111.03634",
    archivePrefix = "arXiv",
    primaryClass = "astro-ph.HE",
    reportNumber = "LIGO-P2100239 ; Data release: https://zenodo.org/record/5655785, LIGO-P2100239",
    doi = "10.1103/PhysRevX.13.011048",
    journal = "Phys. Rev. X",
    volume = "13",
    number = "1",
    pages = "011048",
    year = "2023"
}

@article{Mancarella:2025uat,
    author = "Mancarella, Michele and Gerosa, Davide",
    title = "{Sampling the full hierarchical population posterior distribution in gravitational-wave astronomy}",
    eprint = "2502.12156",
    archivePrefix = "arXiv",
    primaryClass = "gr-qc",
    doi = "10.1103/PhysRevD.111.103012",
    journal = "Phys. Rev. D",
    volume = "111",
    number = "10",
    pages = "103012",
    year = "2025"
}

@misc{gwpopulation_padded,
    title = {Padded GWPopulation},
    author = "Adamcewicz, Christian",
    year = "2026",
    url = "https://github.com/ChristianAdamcewicz/gwpopulation/tree/padded"
}

@misc{gwax,
    title = "{gwax: Gravitational-wave astronomy in JAX}",
    author = "Mould, Matthew",
    year = "2026",
    url = "https://github.com/mdmould/gwax"
}

@misc{flow_repo,
    title = "Normalizing flows for gravitational-wave posteriors",
    author = "Adamcewicz, Christian",
    year = "2026",
    url = "https://github.com/ChristianAdamcewicz/gw_flows"
}

@misc{gwpopulation_pipe_padded,
    title = {Padded GWPopulation\_pipe},
    author = "Adamcewicz, Christian",
    year = "2026",
    url = "https://git.ligo.org/christian.adamcewicz/gwpopulation_pipe/-/tree/padded"}

@article{Dax:2022pxd,
    author = {Dax, Maximilian and Green, Stephen R. and Gair, Jonathan and P{\"u}rrer, Michael and Wildberger, Jonas and Macke, Jakob H. and Buonanno, Alessandra and Sch{\"o}lkopf, Bernhard},
    title = "{Neural Importance Sampling for Rapid and Reliable Gravitational-Wave Inference}",
    eprint = "2210.05686",
    archivePrefix = "arXiv",
    primaryClass = "gr-qc",
    reportNumber = "LIGO-P2200297",
    doi = "10.1103/PhysRevLett.130.171403",
    journal = "Phys. Rev. Lett.",
    volume = "130",
    number = "17",
    pages = "171403",
    year = "2023"
}

@article{Dax:2021tsq,
    author = {Dax, Maximilian and Green, Stephen R. and Gair, Jonathan and Macke, Jakob H. and Buonanno, Alessandra and Sch{\"o}lkopf, Bernhard},
    title = "{Real-Time Gravitational Wave Science with Neural Posterior Estimation}",
    eprint = "2106.12594",
    archivePrefix = "arXiv",
    primaryClass = "gr-qc",
    reportNumber = "LIGO-P2100223",
    doi = "10.1103/PhysRevLett.127.241103",
    journal = "Phys. Rev. Lett.",
    volume = "127",
    number = "24",
    pages = "241103",
    year = "2021"
}

@article{Green:2020dnx,
    author = "Green, Stephen R. and Gair, Jonathan",
    title = "{Complete parameter inference for GW150914 using deep learning}",
    eprint = "2008.03312",
    archivePrefix = "arXiv",
    primaryClass = "astro-ph.IM",
    reportNumber = "LIGO-P2000282",
    doi = "10.1088/2632-2153/abfaed",
    journal = "Mach. Learn. Sci. Tech.",
    volume = "2",
    number = "3",
    pages = "03LT01",
    year = "2021"
}

@article{Talbot:2020oeu,
    author = "Talbot, Colm and Thrane, Eric",
    title = "{Flexible and Accurate Evaluation of Gravitational-wave Malmquist Bias with Machine Learning}",
    eprint = "2012.01317",
    archivePrefix = "arXiv",
    primaryClass = "gr-qc",
    doi = "10.3847/1538-4357/ac4bc0",
    journal = "Astrophys. J.",
    volume = "927",
    number = "1",
    pages = "76",
    year = "2022"
}

@article{Golomb:2021tll,
    author = "Golomb, Jacob and Talbot, Colm",
    title = "{Hierarchical Inference of Binary Neutron Star Mass Distribution and Equation of State with Gravitational Waves}",
    eprint = "2106.15745",
    archivePrefix = "arXiv",
    primaryClass = "astro-ph.HE",
    doi = "10.3847/1538-4357/ac43bc",
    journal = "Astrophys. J.",
    volume = "926",
    number = "1",
    pages = "79",
    year = "2022"
}

@article{Wysocki:2020myz,
    author = "Wysocki, Daniel and O'Shaughnessy, Richard and Wade, Leslie and Lange, Jacob",
    title = "{Inferring the neutron star equation of state simultaneously with the population of merging neutron stars}",
    eprint = "2001.01747",
    archivePrefix = "arXiv",
    primaryClass = "gr-qc",
    reportNumber = "LIGO-P1900359",
    month = "1",
    year = "2020"
}

@article{DEmilio:2021laf,
    author = "D'Emilio, Virginia and Green, Rhys and Raymond, Vivien",
    title = "{Density estimation with Gaussian processes for gravitational wave posteriors}",
    eprint = "2104.05357",
    archivePrefix = "arXiv",
    primaryClass = "gr-qc",
    doi = "10.1093/mnras/stab2623",
    journal = "Mon. Not. Roy. Astron. Soc.",
    volume = "508",
    number = "2",
    pages = "2090--2097",
    year = "2021"
}

@article{Wouters:2025zju,
    author = "Wouters, Thibeau and Pang, Peter T. H. and Koehn, Hauke and Rose, Henrik and Somasundaram, Rahul and Tews, Ingo and Dietrich, Tim and Van Den Broeck, Chris",
    title = "{Leveraging differentiable programming in the inverse problem of neutron stars}",
    eprint = "2504.15893",
    archivePrefix = "arXiv",
    primaryClass = "astro-ph.HE",
    reportNumber = "LA-UR-25-23486",
    doi = "10.1103/v2y8-kxvx",
    journal = "Phys. Rev. D",
    volume = "112",
    number = "4",
    pages = "043037",
    year = "2025"
}

@article{LIGOScientific:2026ctl,
    author = "Abac, None and others",
    collaboration = "LIGO Scientific, VIRGO, KAGRA",
    title = "{GWTC-5.0: Population Properties of Merging Compact Binaries}",
    eprint = "2605.27226",
    archivePrefix = "arXiv",
    primaryClass = "astro-ph.HE",
    reportNumber = "LIGO-P2600045",
    month = "5",
    year = "2026"
}

@article{LIGOScientific:2026wfs,
    author = "Abac, None and others",
    collaboration = "LIGO Scientific, VIRGO, KAGRA",
    title = "{GWTC-5.0: Observations from the Second Part of the Fourth LIGO-Virgo-KAGRA Observing Run and Updates to the Gravitational-Wave Transient Catalog}",
    eprint = "2605.27225",
    archivePrefix = "arXiv",
    primaryClass = "gr-qc",
    reportNumber = "LIGO-P2600152",
    month = "5",
    year = "2026"
}

@article{LIGOScientific:2026ifv,
    author = "Abac, None and others",
    collaboration = "LIGO Scientific, VIRGO, KAGRA",
    title = "{GWTC-5.0: Methods for Identifying and Characterizing Gravitational-wave Transients}",
    eprint = "2605.27224",
    archivePrefix = "arXiv",
    primaryClass = "gr-qc",
    reportNumber = "LIGO-P2600166",
    month = "5",
    year = "2026"
}

@article{LIGOScientific:2026uyd,
    collaboration = "LIGO Scientific, VIRGO, KAGRA",
    title = "{GWTC-5.0: Constraints on the Cosmic Expansion Rate and Modified Gravitational-wave Propagation}",
    eprint = "2605.27227",
    archivePrefix = "arXiv",
    primaryClass = "astro-ph.CO",
    reportNumber = "LIGO-P2600018",
    month = "5",
    year = "2026"
}

@article{LIGOScientific:2026sit,
    author = "Abac, A. G. and others",
    collaboration = "LIGO Scientific, VIRGO, KAGRA",
    title = "{GWTC-5.0: An Introduction to Version 5.0 of the Gravitational-Wave Transient Catalog}",
    eprint = "2605.27223",
    archivePrefix = "arXiv",
    primaryClass = "gr-qc",
    reportNumber = "LIGO-P2500701",
    month = "5",
    year = "2026"
}

@article{LIGOScientific:2026jgl,
    author = "Abac, A. G. and others",
    collaboration = "LIGO Scientific, VIRGO, KAGRA",
    title = "{Open Data from LIGO, Virgo, and KAGRA through the Second Part of the Fourth Observing Run}",
    eprint = "2605.27090",
    archivePrefix = "arXiv",
    primaryClass = "gr-qc",
    reportNumber = "LIGO-P2600085",
    month = "5",
    year = "2026"
}

\end{document}